\documentclass[english,aps,pra,onecolumn,superscriptaddress]{revtex4-1}
\usepackage[latin9]{inputenc}
\usepackage{mathrsfs}
\usepackage{amsmath}
\usepackage{amssymb}
\usepackage{graphicx}

\makeatletter
\usepackage[colorlinks,citecolor=blue,linkcolor=blue]{hyperref}
\usepackage{amsmath}
\usepackage{graphicx}
\allowdisplaybreaks[2]

\makeatother

\usepackage{babel}
\begin{document}
\title{Ultranarrow Superradiant Laser via Dressed Dark State}
\author{Guohui Dong}
\email{dongguohui@sicnu.edu.cn}

\affiliation{College of Physics and Electronic Engineering, Sichuan Normal University,
Chengdu 610068, China}
\author{Yao Yao}
\email{yaoyao_mtrc@caep.cn}

\affiliation{Microsystem and Terahertz Research Center, China Academy of Engineering
Physics, Chengdu 610200, China}
\begin{abstract}
Superradiant laser, which exploits the clock transition of alkaline-earth-metal-like
atoms to generate ultrastable light in the bad-cavity limit, has garnered
much attention in the past few decades. Unlike their odd counterpart,
the even isotopes of alkaline-earth-metal-like atoms possess simpler
structures and longer-lived transitions, which would relax the field
control requirements and enhance the frequency stability of the output
light. However, due to the absence of hyper-fine interaction in even
isotopes, the transition from the state $^{3}\mathrm{P}_{0}$ to $^{1}$S$_{0}$
is strictly forbidden, leading to a vanishing coupling strength between
the cavity mode and the atoms (the state $^{3}\mathrm{P}_{0}$ as
a dark state). In this work, we suggest a superradiant laser scheme
by dressing this dark state with a small bright component by virtue
of a static magnetic field. In contrast to other proposals utilizing
natural atomic transitions, our dressed-state protocol can work from
the crossover regime (coherence in both atoms and photons) to the
superradiant lasing regime (coherence solely in atoms). Specially,
by operating deep into the superradiant lasing regime, our scheme
witnesses a dramatic line-narrowing feature (mHz level) while maintaining
its power. Furthermore, compared to the crossover regime, the laser
frequency in the superradiant lasing regime is more robust against
the fluctuations of the cavity length and magnetic field strength.
Our proposal demonstrates the potential for extracting ultranarrow
light from even isotopes of the alkaline-earth-metal-like atoms and
may find its role in frequency stabilization and related precision
measurement scenarios.
\end{abstract}
\maketitle

\section{Introduction}

Since the first detection in 1960 \citep{MAIMAN1960}, lasers, especially
frequency-stabilized representatives, have found applications in diverse
areas, such as optical atomic clocks \citep{Derevianko2011,Ludlow2015}
and atom or ion cooling \citep{atomcooling1975,ioncooling1975,Cohen_Tannoudji1990}.
Typically, the laser frequency is stabilized using feedback from a
high-finesse optical cavity whose length is ultimately limited by
the thermal noise of the end mirrors \citep{Numata2004}. Consequently,
significant efforts have been devoted to reducing the Brownian noise
of the mirrors utilizing high-quality materials, for instance, ultralow
expansion (ULE) glass \citep{Numata2004,Yu2019}, sapphire \citep{Storz1998},
and single-crystal silicon \citep{Kessler2012}. Recently, as an alternative
route, a superradiant laser has gained considerable attention due
to its potential to extract light directly from atomic clock transitions.
Specifically, in the bad-cavity limit where the photon dissipation
rate is several orders of magnitude larger than the atomic relaxation
rate, the atoms behave as a macroscopic spin through interacting with
the same cavity mode \citep{Meiser2009,Meiser2010,Meiser2010a,Liu2020}.
Therefore, above a certain pump, the atoms emit light collectively,
resembling the transient superradiance, but in a steady-state manner.
As the atomic phases are synchronized, the superradiant emission light
is line-narrowed, even narrower than the natural linewidth of the
atomic transition. Unlike the conventional laser where the coherence
is mainly stored in photons, the coherence of the superradiant laser
is significantly influenced by the macroscopic atomic spin, demonstrating
strong atom-atom correlations. Particularly, the superradiant laser
can operate in two regimes: the crossover regime, where both the stimulated
emission photons and macroscopic spin contribute to the system's coherence,
and the superradiant lasing regime, where in contrast the coherence
is solely in atoms \citep{Norcia2016,Holland2017,Debnath2018}.

The superradiant laser requires a long-lifetime atomic clock transition,
rendering the intercombination transitions in the alkaline-earth-metal-like
atoms (e.g., Sr, Ca, Mg, Yb) favorable candidates. For example, leveraging
the $7.5$kHz spin-forbidden $^{3}$P$_{1}$-$^{1}$S$_{0}$ transition
of the $^{88}$Sr atoms, the Ref. \citep{Norcia2016} achieves quasi-steady-state
emission with a linewidth of around 6kHz. For a system composed of
the $^{87}$Sr atoms ($I=9/2$), the weakly allowed double-forbidden
transition $^{3}$P$_{0}$-$^{1}$S$_{0}$ (linewidth around mHz)
yields a millihertz superradiant laser \citep{Meiser2009}. Further
pursuit of the stabilized laser can hinge on atomic transitions with
longer lifetimes, such as the $^{3}$P$_{0}$-$^{1}$S$_{0}$ transition
in even isotopes of the alkaline-earth-metal-like atoms ($I=0$, lifetime
around several thousands of years) \citep{Santra2004}. Additionally,
compared with their odd counterpart with hyperfine interaction, the
simpler structure of even isotopes exhibits scalar polarizability,
thereby easing the field control requirement in the lasing scheme
\citep{Santra2005,Barber2006}. However, the absence of the nuclear
spin also indicates a vanishing single-photon transition matrix element
($^{3}$P$_{0}$ acting as a dark state), resulting in a negligible
photon number in the steady state. In this sense, a proposal which
can effectively couple the dark-state transition of even alkaline-earth-metal-like
atoms to the cavity mode is urgently desired for the ultranarrow superradiant
laser.

One approach for achieving this effective interaction is utilizing
an intermediate level via external fields \citep{Santra2005,Dong2023,Barber2006,Taichenachev2006,Dubey2025}.
For instance, although the transition from the clock state $^{3}$P$_{0}$
to the ground state $^{1}$S$_{0}$ is strictly forbidden, it can
be mediated by the broad state $^{1}$P$_{1}$ with probe and dressing
lasers \citep{Santra2005}. In our previous proposal of coherence-assisted
superradiant laser, we couple the atomic ground state $^{1}$S$_{0}$
with the clock state $^{3}$P$_{0}$ via two Raman beams, where the
frequencies of the beams demand judicious control \citep{Dong2023}.
Inspired by the work of magnetic-induced excitation of forbidden transition
\citep{Barber2006,Taichenachev2006,Dubey2025}, here we explore the
dark state superradiant laser system with a static magnetic field.
In our proposal, the dark state $^{3}$P$_{0}$ is dressed with a
small bright $^{3}$P$_{1}$ component via the magnetic dipole (M1)
transition, and hence photons will be emitted when the cavity mode
is resonant with this dressed dark state. Originating from the M1
transition, the lasing properties of our protocol, such as the threshold,
steady-state radiation or power, and the coherence feature of the
laser, can be manipulated by the magnetic field. Therefore, in contrast
to other proposals utilizing natural atomic transitions, our dressed-state
scenario can operate from the crossover regime (coherence in both
atoms and photons) to the superradiant lasing regime (coherence solely
in atoms). Specially, by operating deep into the superradiant lasing
regime, our superradiant laser scheme witnesses a dramatic line-narrowing
feature (mHz level) while maintaining its power. Furthermore, compared
to the crossover regime, the laser frequency in the superradiant lasing
regime is more robust against the fluctuations of the cavity length
and magnetic field strength. Our magnetic-field-tuned proposal provides
an alternative route for the superradiant laser system utilizing even
alkaline-earth-metal-like atoms, and may find potential applications
in frequency stabilization and the related precision measurement scenarios.

\section{Model}

\subsection{System Setup}

\begin{figure}[t]
\begin{centering}
\includegraphics[scale=0.5]{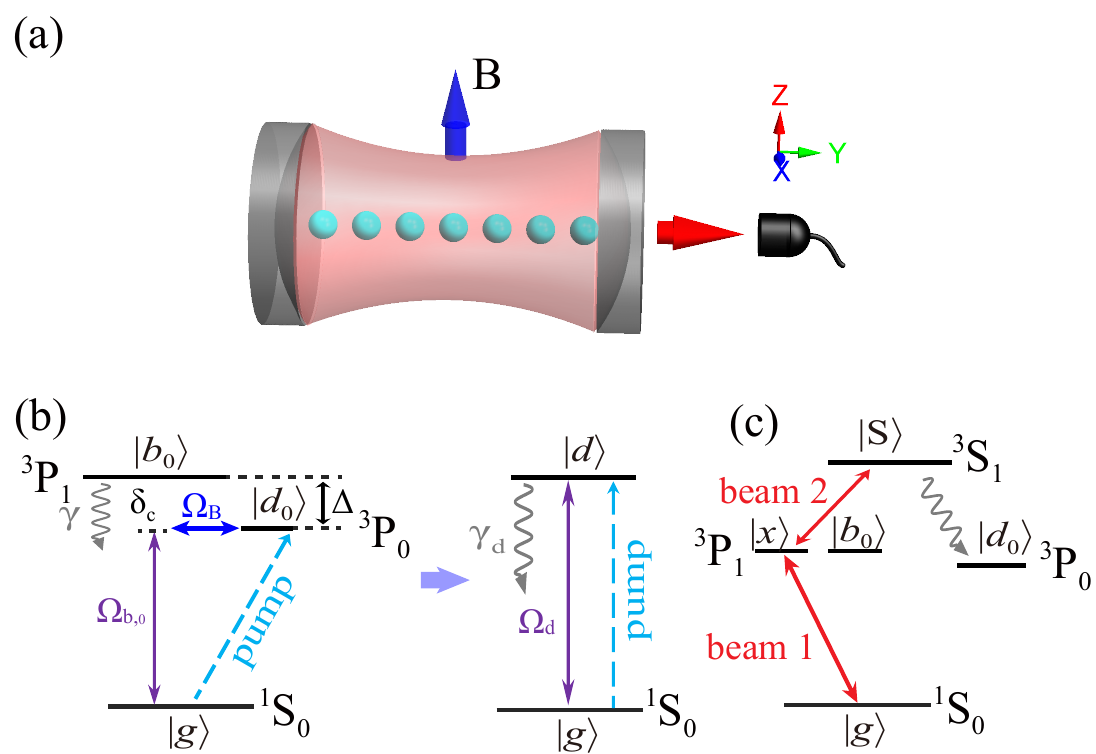}
\par\end{centering}
\caption{\label{fig:schematic diagram}(a) $N$ cooled bosonic alkaline-earth-metal-like
atoms are trapped in the cavity where a static magnetic field is applied.
(b) Schematic diagrams of the three-level and effective two-level
models. The cavity mode couples with the bright transition $^{3}$P$_{1}$-$^{1}$S$_{0}$.
A static magnetic field interacts with $^{3}$P$_{0}$ and $^{3}$P$_{1}$
states through the M1 transition. The atoms are pumped incoherently
from the ground state $^{1}$S$_{0}$ to the dark state $^{3}$P$_{0}$.
(c) One possible pump process with two pumping beams.}
\end{figure}
In our scenario, $N$ cold bosonic alkaline-earth-metal-like atoms
are confined in an optical cavity whose axis is along the $y$-direction
(see Fig. \ref{fig:schematic diagram}(a)). The cavity mode couples
with the intercombination transition of the atoms, i.e., the ground
state $\left|g\right\rangle \equiv\left|^{1}\mathrm{S}_{0}\right\rangle $
and the sublevel of the bright state $\left|b_{0}\right\rangle \equiv\left|^{3}\mathrm{P}_{1},m=0\right\rangle $
where $m$ is the magnetic quantum number. Due to the absence of the
nuclear spin, the $^{3}$P$_{0}$-$^{1}$S$_{0}$ transition in even
isotopes is strictly forbidden, leaving the state $\left|d_{0}\right\rangle \equiv\left|^{3}\mathrm{P}_{0}\right\rangle $
as a dark state (no lasing). Actually, a $z$-directioned static magnetic
field would couple the dark state with the bright one through the
M1 transition and thus induce an intermediate coupling between the
cavity mode and the dark transition. The Hamiltonian of the system
reads ($\hbar=1$)
\begin{align}
\hat{H} & =\hat{H}_{a}+\hat{H}_{c}+\hat{H}_{a-c},\label{eq:total Hamil}\\
\hat{H}_{a} & =\sum_{i=1}^{N}(\omega_{b,0}\left|b_{0}\right\rangle _{i}\left\langle b_{0}\right|+\omega_{d,0}\left|d_{0}\right\rangle _{i}\left\langle d_{0}\right|)+\sum_{i=1}^{N}\left(\Omega_{B}\left|b_{0}\right\rangle _{i}\left\langle d_{0}\right|+h.c.\right),\\
\hat{H}_{c} & =\omega_{c}\hat{c}^{\dagger}\hat{c},\\
\hat{H}_{a-c} & =\sum_{i=1}^{N}\left(\frac{\Omega_{b,0}}{2}\left|b_{0}\right\rangle _{i}\left\langle g\right|\hat{c}+h.c.\right),
\end{align}
where $\hat{c}$ $(\hat{c}^{\dagger})$ represents the annihilation
(creation) operator of the cavity mode with frequency $\omega_{c}$.
$\omega_{b,0}$ ($\omega_{d,0}$) is the frequency of the atomic bare
bright (dark) state which couples with the cavity mode with the Rabi
frequency $\Omega_{b,0}$ ($\Omega_{d,0}=0$). A $z$-directioned
magnetic field $B$ interacts with the bright-dark transition via
the M1 process with the Rabi frequency $\Omega_{B}$. For alkaline-earth-metal-like
atoms, the magnetic Rabi frequency is $\Omega_{B}=2\pi\times1.14B$MHz
($B$ given in Gaussian unit) \citep{Taichenachev2006}. The frequency
gap $\Delta\equiv\omega_{b,0}-\omega_{d,0}$ between the atomic bare
bright and dark states is around $2\pi\times1$THz \citep{Courtillot2005}.

As shown in Fig. \ref{fig:schematic diagram}(b), the atoms are pumped
from the ground state $\left|g\right\rangle $ to the bare-dark state
$\left|d_{0}\right\rangle $ incoherently. This pumping process can
be achieved by using several beams to coherently pump the atom from
$\left|g\right\rangle $ to an auxiliary excited state which then
decays rapidly to $\left|d_{0}\right\rangle $ via spontaneous emission.
A typical example is illustrated in Fig. \ref{fig:schematic diagram}(c).
An $x$-polarized beam 1 couples $\left|g\right\rangle $ with the
sublevel of $^{3}$P$_{1}$, i.e., $\left|x\right\rangle \equiv\left(\left|^{3}\mathrm{P}_{1},m=-1\right\rangle -\left|^{3}\mathrm{P}_{1},m=1\right\rangle \right)/\sqrt{2}$
while a $y$-polarized beam 2 couples $\left|x\right\rangle $ with
a sublevel of $^{3}$S$_{1}$ ($\left|\mathrm{S}\right\rangle \equiv\left|^{3}\mathrm{S}_{1},m=0\right\rangle $).
Finally, atoms in $\left|\mathrm{S}\right\rangle $ decay to $\left|d_{0}\right\rangle $
at a rate around several MHz. Notice that the transition from $\left|\mathrm{S}\right\rangle $
to $\left|b_{0}\right\rangle $ is dipole forbidden, i.e., no pump
to $\left|b_{0}\right\rangle $.

\subsection{Lasing from the dressed dark state}

Although the cavity mode couples with the bright transition of the
atoms, the vanishing pump indicates no population in the $^{3}$P$_{1}$
state and thus no lasing. In fact, since the atoms are continuously
pumped to the $^{3}$P$_{0}$ state, in the steady state, the atoms
will all be trapped in this dark state.

In the presence of the static magnetic field, the bright state also
interacts with the dark state, which effectively opens the $^{3}$P$_{0}$-
$^{1}$S$_{0}$ dark transition. In fact, this magnetic-field-induced
lasing process can be elucidated more evidently with the dressed-state
basis. Actually, for a moderate magnetic field ($B$ no more than
$1\mathrm{T}=10^{4}\mathrm{G}$), the magnetic Rabi frequency $\Omega_{B}$
is much smaller than the frequency gap of the bright-dark transition
(THz level). Hence, by treating the M1 transition term as a perturbation,
the Hamiltonian in the dressed-state basis can be recast as 
\begin{align}
\hat{H} & =\omega_{c}\hat{c}^{\dagger}\hat{c}+\sum_{i=1}^{N}\left(\omega_{b}\left|b\right\rangle _{i}\left\langle b\right|+\omega_{d}\left|d\right\rangle _{i}\left\langle d\right|\right)+\sum_{i=1}^{N}\left(\frac{\Omega_{b}}{2}\left|b\right\rangle _{i}\left\langle g\right|\hat{c}+\frac{\Omega_{d}}{2}\left|d\right\rangle _{i}\left\langle g\right|\hat{c}+h.c.\right),\label{eq:3-level dressed state Hamil}
\end{align}
where $\left|b\right\rangle \simeq\left|b_{0}\right\rangle +\Omega_{B}/\Delta\left|d_{0}\right\rangle $
($\left|d\right\rangle \simeq\left|d_{0}\right\rangle -\Omega_{B}/\Delta\left|b_{0}\right\rangle $)
stands for the dressed-bright (dressed-dark) state involving a small
dark (bright) component. The Zeeman-shifted frequencies of the two
dressed states are $\omega_{b}=\omega_{b,0}+\varpi_{B}$ and $\omega_{d}=\omega_{d,0}-\varpi_{B}$
with $\varpi_{B}\equiv-\Delta/2+\sqrt{(\Delta/2)^{2}+\Omega_{B}^{2}}$.
The dressed-bright (dressed-dark) state coherently interacts with
the cavity mode with the Rabi frequency $\Omega_{b}\simeq\Omega_{b,0}$
($\Omega_{d}\simeq\Omega_{b,0}\Omega_{B}/\Delta$). Moreover, as the
dark component in $\left|b\right\rangle $ is rather small, the pump
to this dressed-bright state is still negligible, i.e., the atomic
population in $\left|b\right\rangle $ is vanishing (see the following
discussions). Accordingly, one can simplify this scheme to a two-level
model by ignoring the dressed-bright state in Eq. (\ref{eq:3-level dressed state Hamil}).
The approximated two-level dressed-dark-state Hamiltonian can be rewritten
as (in the interaction picture)
\begin{align}
\hat{H}_{I} & =\delta_{c}\hat{c}^{\dagger}\hat{c}+\sum_{i=1}^{N}\left(\frac{\Omega_{d}}{2}\left|d\right\rangle _{i}\left\langle g\right|\hat{c}+h.c.\right),\label{eq:2-level dressed state Hamil}
\end{align}
where $\delta_{c}\equiv\omega_{c}-\omega_{d}$ denotes the cavity-atom
detuning (see Fig. \ref{fig:schematic diagram}(b)). In consequence,
with the magnetic field, the long-lived dark state is dressed with
a small bright component and thus coherently couples with the cavity
mode. Above a certain threshold, the atomic population inversion would
be generated between the ground and dressed-dark states, leading to
the output of lasing photons.

\section{Dynamical Equations and Analysis}

\subsection{Dynamical Equations}

In the effective two-level model, with the incoherent pump and dissipation
processes, the dynamics of the density matrix $\hat{\rho}\equiv\hat{\rho}(t)$
of the total system is governed by the following master equation 
\begin{align}
\frac{d\hat{\rho}}{dt} & =-i[\hat{H}_{I},\hat{\rho}]+\kappa\mathscr{L}[\hat{c}]\hat{\rho}+\sum_{i=1}^{N}\left(\eta\mathscr{L}[\hat{\sigma}_{i}^{\dagger}]+\gamma_{d}\mathscr{L}[\hat{\sigma}_{i}]\right)\hat{\rho},\label{eq:master euqation}
\end{align}
where the Lindblad operator is defined as $\mathscr{L}[\hat{\mathscr{O}}]\equiv\hat{\mathscr{O}}\hat{\rho}\hat{\mathscr{O}}^{\dagger}-\hat{\mathscr{O}}^{\dagger}\hat{\mathscr{O}}\hat{\rho}/2-\hat{\rho}\hat{\mathscr{O}}^{\dagger}\hat{\mathscr{O}}/2$.
$\kappa$ ($\eta$, $\gamma_{d}\equiv\gamma\Omega_{B}^{2}/\Delta^{2}$)
denotes the dissipation rate of the photon (pump rate, decay rate
of the atom). $\hat{\sigma}_{i}\equiv\left|g\right\rangle _{i}\left\langle d\right|$
stands for the transition operator of the $i$th atom.

Owing to the infinite dimension of the cavity mode and the exponential
scaling of the dimension of gain medium with the atomic number, approximate
methods for solving the above master equation are urgently desired.
Although the first-order mean-field theory yields useful results,
for instance, the steady-state photon number and atomic inversion,
it violates the phase symmetry and overlooks the quantum correlations
of the system which are crucial for characterizing the coherence of
the system. Here, beyond the first-order method, we solve the dynamical
equations of the system within the second-order mean-field theory
where the correlation between any two operators is retained and the
phase symmetry of the system is preserved \citep{Meiser2009,Dong2023}.

In the second-order mean-field theory, the correlation of three or
more operators is ignored, i.e., 
\begin{align}
\left\langle \hat{X}\hat{Y}\hat{Z}\right\rangle  & \simeq\left\langle \hat{X}\right\rangle \left\langle \hat{Y}\hat{Z}\right\rangle +\left\langle \hat{Y}\right\rangle \left\langle \hat{X}\hat{Z}\right\rangle +\left\langle \hat{Z}\right\rangle \left\langle \hat{X}\hat{Y}\right\rangle -2\left\langle \hat{X}\right\rangle \left\langle \hat{Y}\right\rangle \left\langle \hat{Z}\right\rangle .
\end{align}
Here $\left\langle \hat{X}\right\rangle \equiv\left\langle \hat{X}(t)\right\rangle \equiv\mathrm{Tr}(\hat{X}\hat{\rho})$
represents the expectation value of the operator $\hat{X}$. It has
been demonstrated that the density matrix $\hat{\rho}$ is symmetric
under the phase transformation $U(\theta)\equiv\exp[i\theta(\hat{c}^{\dagger}\hat{c}+\sum_{j=1}^{N}\hat{\sigma}_{j}^{\dagger}\hat{\sigma}_{j})]$
for any $\theta$. As a result, the expectation values of operators
which are not symmetric would vanish. Moreover, as all the atoms interact
with the field homogeneously, we can simplify the expressions of expectation
values as $\left\langle \hat{\sigma}_{z,i}\right\rangle =\left\langle \hat{\sigma}_{z}\right\rangle \equiv\left\langle \hat{\sigma}^{\dagger}\hat{\sigma}-\hat{\sigma}\hat{\sigma}^{\dagger}\right\rangle $,
$\left\langle \hat{c}^{\dagger}\hat{\sigma}_{i}\right\rangle \equiv\left\langle \hat{c}^{\dagger}\hat{\sigma}\right\rangle $,
and $\left\langle \hat{\sigma}_{i}^{\dagger}\hat{\sigma}_{j\neq i}\right\rangle =\left\langle \hat{\sigma}_{1}^{\dagger}\hat{\sigma}_{2}\right\rangle $
for all $i$.

As a first step, the dynamical equation for the expectation value
of the photon number reads

\begin{equation}
\frac{d}{dt}\left\langle \hat{c}^{\dagger}\hat{c}\right\rangle =-\kappa\left\langle \hat{c}^{\dagger}\hat{c}\right\rangle +\frac{N\Omega_{d}}{2i}\left(\left\langle \hat{c}^{\dagger}\hat{\sigma}\right\rangle -\left\langle \hat{\sigma}^{\dagger}\hat{c}\right\rangle \right).\label{eq:photon number}
\end{equation}
 Apparently, the dynamics of the photon number is related to the atom-field
correlation $\left\langle \hat{c}^{\dagger}\hat{\sigma}\right\rangle $.
To close the dynamical equation, we obtain the derivatives of the
atomic inversion $\left\langle \hat{\sigma}_{z}\right\rangle $, the
atom-atom correlation $\left\langle \hat{\sigma}_{1}^{\dagger}\hat{\sigma}_{2}\right\rangle $,
and the atom-field correlation $\left\langle \hat{c}^{\dagger}\hat{\sigma}\right\rangle $
as follows
\begin{align}
\frac{d}{dt}\left\langle \hat{\sigma}_{z}\right\rangle  & =-\Gamma\left(\left\langle \hat{\sigma}_{z}\right\rangle -d_{0}\right)+i\Omega_{d}\left(\left\langle \hat{c}^{\dagger}\hat{\sigma}\right\rangle -\left\langle \hat{\sigma}^{\dagger}\hat{c}\right\rangle \right),\label{eq:inversion}\\
\frac{d}{dt}\left\langle \hat{\sigma}_{1}^{\dagger}\hat{\sigma}_{2}\right\rangle  & =-\Gamma\left\langle \hat{\sigma}_{1}^{\dagger}\hat{\sigma}_{2}\right\rangle +\frac{\Omega_{d}\left\langle \hat{\sigma}_{z}\right\rangle }{2i}\left(\left\langle \hat{c}^{\dagger}\hat{\sigma}\right\rangle -\left\langle \hat{\sigma}^{\dagger}\hat{c}\right\rangle \right),\label{eq:atom-atom corr}\\
\frac{d}{dt}\left\langle \hat{c}^{\dagger}\hat{\sigma}\right\rangle  & =-\left(\frac{\Gamma+\kappa}{2}-i\delta_{c}\right)\left\langle \hat{c}^{\dagger}\hat{\sigma}\right\rangle +i\frac{\Omega_{d}}{2}\frac{\left\langle \hat{\sigma}_{z}\right\rangle +1}{2}+i\frac{\Omega_{d}}{2}\left(\left\langle \hat{c}^{\dagger}\hat{c}\right\rangle \left\langle \hat{\sigma}_{z}\right\rangle +(N-1)\left\langle \hat{\sigma}_{1}^{\dagger}\hat{\sigma}_{2}\right\rangle \right),\label{eq:atom-field corr}
\end{align}
where we have defined the notations $\Gamma\equiv\eta+\gamma_{d}$
and $d_{0}\equiv(\eta-\gamma_{d})/\Gamma$. To demonstrate the validity
of the two-level dressed-dark-state model, we also present the dynamical
equations of the original three-level model (see Appendix \ref{sec:Three level equation}).

\begin{table}[t]
\begin{centering}
\includegraphics[scale=0.35]{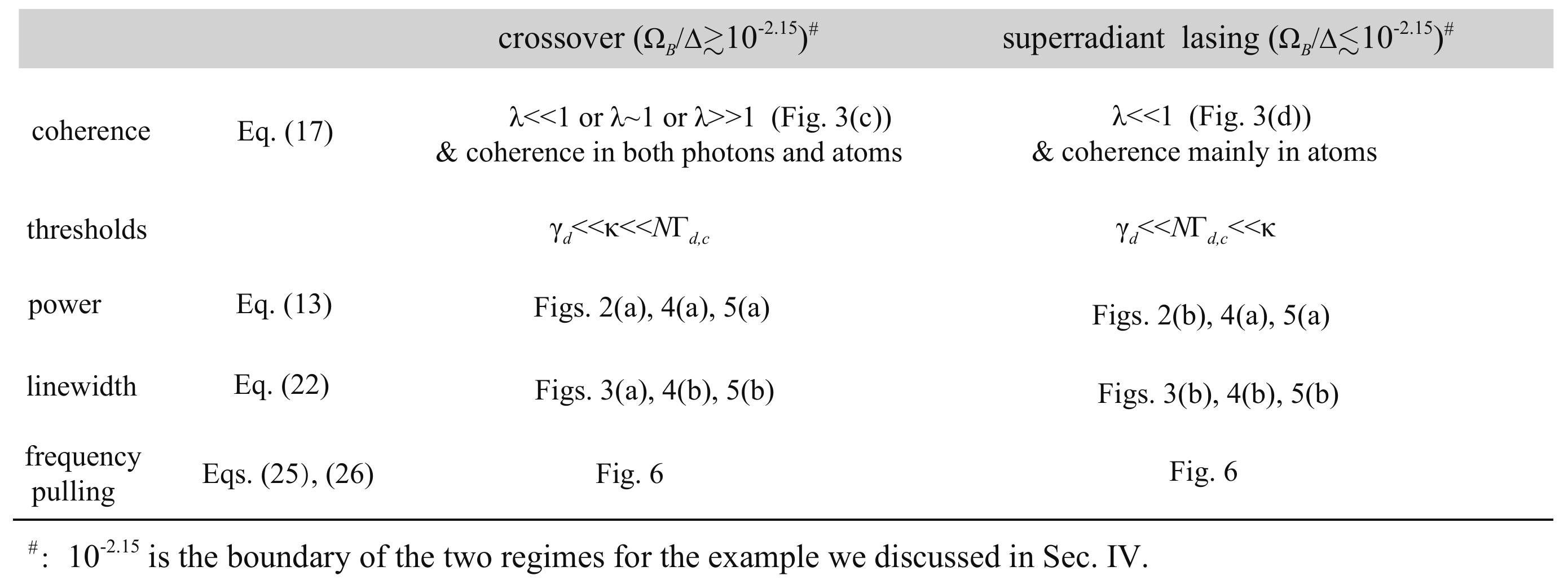}
\par\end{centering}
\caption{\label{tab:1}The laser features in the crossover and superradiant
lasing regimes.}
\end{table}

In the bad-cavity limit, it has been elucidated that the atoms are
collectively correlated in phase in the lasing region \citep{Meiser2009,Meiser2010,Debnath2018}.
Thus for a large atomic number $N\gg1$, the collective correlation
$N\left\langle \hat{\sigma}_{1}^{\dagger}\hat{\sigma}_{2}\right\rangle $
in Eq. (\ref{eq:atom-field corr}) would be much larger than the single-atom
population inversion $\left\langle \hat{\sigma}_{z}\right\rangle $.
Hence, by neglecting the small term $i\Omega_{d}\left[\left(\left\langle \hat{\sigma}_{z}\right\rangle +1\right)/2-\left\langle \hat{\sigma}_{1}^{\dagger}\hat{\sigma}_{2}\right\rangle \right]/2$
in Eq. (\ref{eq:atom-field corr}), one can achieve the approximate
analytical steady-state solutions of the above dynamical equations
(for the resonant case $\delta_{c}=0$)
\begin{align}
\left\langle \hat{c}^{\dagger}\hat{c}\right\rangle _{\mathrm{s}}^{(0)} & =\frac{N^{2}\Omega_{d}^{2}}{\kappa^{2}}\frac{d_{0}-\frac{1}{C}}{2C},\label{eq:appro solu-1}\\
\left\langle \hat{\sigma}_{z}\right\rangle _{\mathrm{s}}^{(0)} & =\frac{1}{C},\label{eq:appro solu-2}\\
\left\langle \hat{\sigma}_{1}^{\dagger}\hat{\sigma}_{2}\right\rangle _{\mathrm{s}}^{(0)} & =\frac{d_{0}-\frac{1}{C}}{2C},\label{eq:appro solu-3}\\
\left\langle \hat{c}^{\dagger}\hat{\sigma}\right\rangle _{\mathrm{s}}^{(0)} & =i\frac{N\Omega_{d}}{\kappa}\frac{d_{0}-\frac{1}{C}}{2C},\label{eq:appro solu-4}
\end{align}
where $C=N\Omega_{d}^{2}/\kappa\Gamma$ is the generalized collective
cooperative parameter. Here the subscript ``$\mathrm{s}$'' stands
for the steady state while the superscript ``$(0)$'' indicates
that Eqs. (\ref{eq:appro solu-1})-(\ref{eq:appro solu-4}) is the
lowest-order approximate results. The high-order corrections of the
steady-state solution can be retained by taking $i\Omega_{d}[(\left\langle \hat{\sigma}_{z}\right\rangle +1)/2-\left\langle \hat{\sigma}_{1}^{\dagger}\hat{\sigma}_{2}\right\rangle ]/2$
in Eq. (\ref{eq:atom-field corr}) as a perturbation (see Appendix
\ref{sec:appro analy solu}).

It is worth to mention that Eqs. (\ref{eq:photon number})-(\ref{eq:atom-field corr})
demonstrate the significance of coherence in the laser system. As
is indicated, the growth of the intracavity photon number and atomic
inversion relies on the field-atom coherence $\left\langle \hat{c}^{\dagger}\hat{\sigma}\right\rangle $
which is inherited from $\left\langle \hat{c}^{\dagger}\hat{c}\right\rangle \left\langle \hat{\sigma}_{z}\right\rangle $
and $N\left\langle \hat{\sigma}_{1}^{\dagger}\hat{\sigma}_{2}\right\rangle $.
Roughly speaking, $\left\langle \hat{c}^{\dagger}\hat{c}\right\rangle \left\langle \hat{\sigma}_{z}\right\rangle $
characterizes the strength of the stimulation emission while $N\left\langle \hat{\sigma}_{1}^{\dagger}\hat{\sigma}_{2}\right\rangle $
represents the atom-atom correlation in the collective spin. In this
sense, one can quantify the relative contribution of photons and atoms
to the coherence of the system with the coherence contribution parameter

\begin{equation}
\lambda\equiv\frac{\left\langle \hat{c}^{\dagger}\hat{c}\right\rangle \left\langle \hat{\sigma}_{z}\right\rangle }{N\left\langle \hat{\sigma}_{1}^{\dagger}\hat{\sigma}_{2}\right\rangle }.\label{eq:coh-contri-para}
\end{equation}
With the approximate analytical solution in Eqs. (\ref{eq:appro solu-1})-(\ref{eq:appro solu-3}),
the coherence contribution parameter can be simplified as the ratio
of the effective dissipation rate of the atom $\Gamma$ over the photon
decay rate $\kappa$, i.e., $\lambda\simeq\Gamma/\kappa$. Obviously,
$\lambda\gg1$ ($\lambda\ll1$) implies the coherence mainly in photons
(atoms) while $\lambda\sim1$ corresponds to the compatible case where
the system coherence is in both photons and atoms. In fact, the coherence
feature of the laser varies with system parameters, such as the magnetic
field strength and pump rate. For reasons that would be clear in the
next section, we refer to the system as operating in the crossover
regime when the coherence can be stored in both atoms and photons,
i.e., the coherence contribution parameter varies from $\lambda\ll1$
to $\lambda\gg1$ in the lasing region, while the superradiant lasing
regime when the coherence is kept solely in atoms, i.e., $\lambda\ll1$
throughout the lasing region. In our scenario, the laser system can
operate from the crossover regime to the superradiant lasing regime
with the modulation of the magnetic field, where the laser properties
are distinct from one another (see Tab. \ref{tab:1} and the following
discussion for more details).

\begin{figure}[t]
\begin{centering}
\includegraphics[scale=0.3]{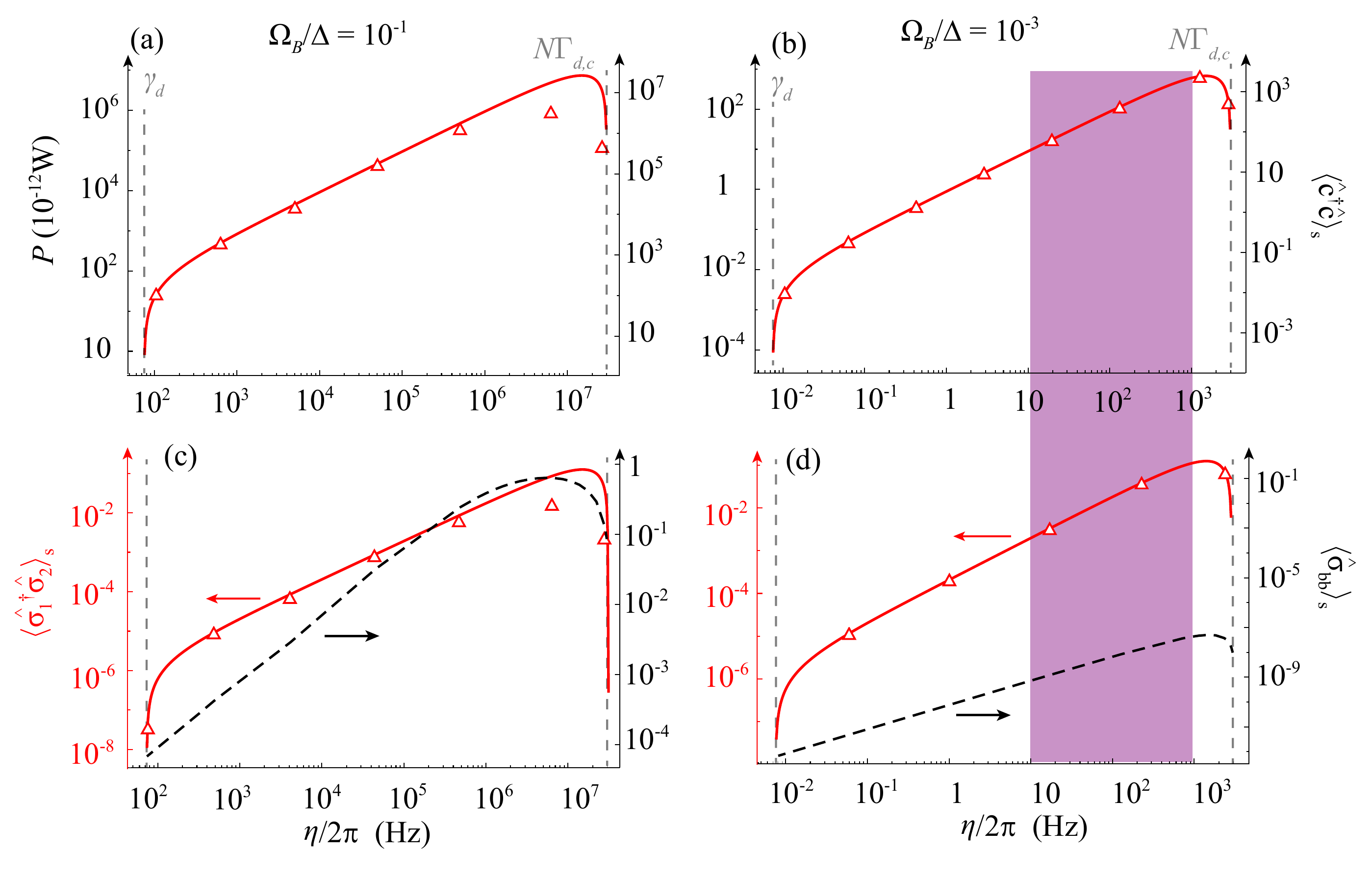}
\par\end{centering}
\caption{\label{fig:2}The output power $P=\kappa\hbar\omega\left\langle \hat{c}^{\dagger}\hat{c}\right\rangle _{\mathrm{s}}$,
steady-state photon number $\left\langle \hat{c}^{\dagger}\hat{c}\right\rangle _{\mathrm{s}}$,
atom-atom correlation $\left\langle \hat{\sigma}_{1}^{\dagger}\hat{\sigma}_{2}\right\rangle $,
and dressed-bright state population $\left\langle \hat{\sigma}_{bb}\right\rangle _{\mathrm{s}}\equiv\lim_{t\rightarrow\infty}\mathrm{Tr}[\left|b\right\rangle \left\langle b\right|\hat{\rho}(t)]$
as a function of the pump rate $\eta$ for $\Omega_{B}/\Delta=10^{-1}$
(the crossover regime) (a,c) and $\Omega_{B}/\Delta=10^{-3}$ (the
superradiant lasing regime) (b,d). The gray dashed vertical lines
denote two thresholds of the laser, i.e. $\gamma_{d}\equiv\gamma\Omega_{B}^{2}/\Delta^{2}$
and $N\Gamma_{d,c}\equiv N\Omega_{b,0}^{2}\Omega_{B}^{2}/(\kappa\Delta^{2})$.
The coincidence of the two-level approximate analytical (red solid
lines) and three-level numerical results (red triangles) elucidates
the effectiveness of the two-level model which can be ascribed to
the negligible population of the dressed-bright state (black dashed
lines). The region satisfying $P\protect\geq10^{-11}\mathrm{W}$,
$\Delta\nu\protect\leq2\pi\times10\mathrm{mHz}$, $\eta\protect\leq2\pi\times10^{3}\mathrm{Hz}$
is marked in purple.}
\end{figure}

\subsection{Laser spectrum\label{subsec:Laser-spectrum}}

Beyond the photon number in the steady state, the laser linewidth
is another significant feature of the output laser. As the full width
at half maximum (FWHM) of the laser spectrum, the laser linewidth
describes the spread of the laser around its central frequency and
characterizes the spectral purity of the laser. According to the Wiener-Kintchine
theorem \citep{Wiener1930,Khintchine1934,Huang2009}, the laser spectrum
is related to the correlation function of the cavity mode as
\begin{equation}
S(\omega)=2\int_{0}^{\infty}dt\mathrm{Re}\left[\left\langle \hat{c}^{\dagger}(t)\hat{c}(0)\right\rangle _{\mathrm{s}}e^{-i(\omega-\omega_{d})t}\right].
\end{equation}
\begin{figure}[t]
\begin{centering}
\includegraphics[scale=0.3]{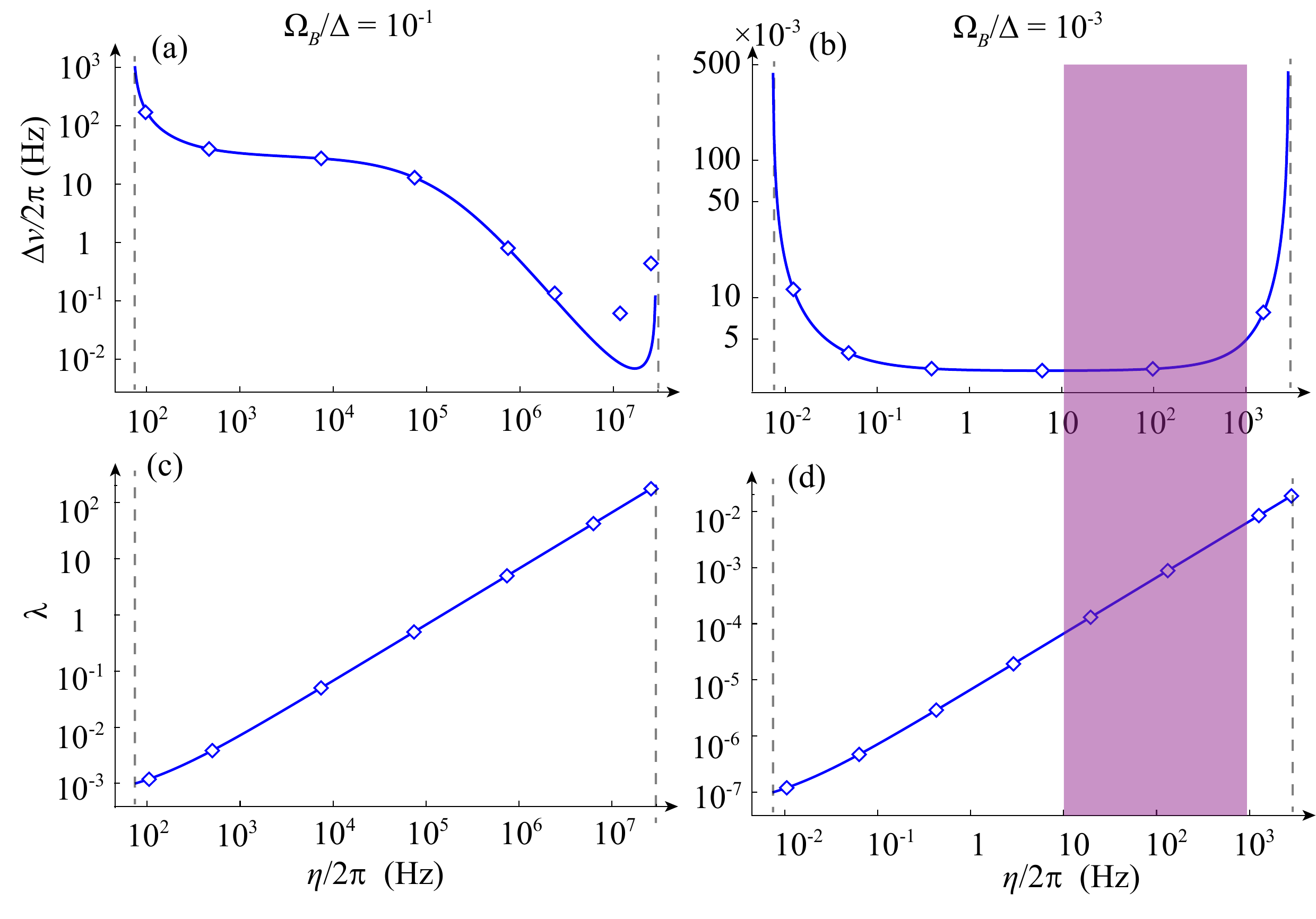}
\par\end{centering}
\caption{\label{fig:3}The linewidth $\Delta\nu$ (a,b) and coherence contribution
parameter $\lambda$ (c,d) of the superradiant laser as a function
of the pump rate $\eta$. The linewidth in the $\Omega_{B}/\Delta=10^{-1}$
case (the crossover regime, the coherence contribution parameter from
$\lambda\ll1$ to $\lambda\gg1$) firstly displays a constant character
and then turns to a Schawlow--Townes-like feature, while that in
the $\Omega_{B}/\Delta=10^{-3}$ case (the superradiant lasing regime,
$\lambda\ll1$) remains almost constant. The blue solid lines (diamonds)
plot the approximate analytical results of the two-level dressed-dark-state
model (the numerical results of the original three-level model). The
region satisfying $P\protect\geq10^{-11}\mathrm{W}$, $\Delta\nu\protect\leq2\pi\times10\mathrm{mHz}$,
and $\eta\protect\leq2\pi\times10^{3}\mathrm{Hz}$ is marked in purple.}
\end{figure}
Within the frame of the second-order mean-field theory, the dynamical
equation of the correlation function $\left\langle \hat{c}^{\dagger}(t)\hat{c}(0)\right\rangle _{\mathrm{s}}$
can be cast as
\begin{equation}
\frac{d}{dt}V(t)=\mathbb{W}V(t),\label{eq:quan regress}
\end{equation}
where we have defined the vector $V(t)\equiv\left(\left\langle \hat{c}^{\dagger}(t)\hat{c}(0)\right\rangle _{\mathrm{s}},\left\langle \hat{\sigma}^{\dagger}(t)\hat{c}(0)\right\rangle _{\mathrm{s}}\right)^{T}$
and matrix 
\begin{equation}
\mathbb{W}=\frac{1}{2}\left(\begin{array}{cc}
2i\delta_{c}-\kappa & iN\Omega_{d}\\
-i\Omega_{d}\left\langle \hat{\sigma}_{z}\right\rangle _{\mathrm{s}} & -\Gamma
\end{array}\right).
\end{equation}
The correlation function vector $V(t)$ can be expressed as \citep{Dong2023}
\begin{align}
V(t) & =e^{\mathbb{W}t}V(0)=\sum_{j=1}^{2}e^{w_{j}t}\left|j\right\rangle \left\langle \widetilde{j}\right|V(0),
\end{align}
where $w_{j}$ is the $j$th eigenvalue of $\mathbb{W}$ with the
right and left eigenvectors $\left|j\right\rangle $ and $\left\langle \widetilde{j}\right|$
for $i,j=1,2$. The initial condition $V(0)$ corresponds to the steady-state
solutions of Eqs. (\ref{eq:photon number}) and (\ref{eq:atom-field corr}).
Therefore, the laser spectrum behaves as a weighted superposition
of two Lorentzian lineshapes. Concretely, the $j$th lineshape is
centered around the frequency $\omega_{d}+\mathrm{Im}(w_{j})$ with
linewidth $2\left|\mathrm{Re}(w_{j})\right|$. As illustrated in the
experimental and theoretical works \citep{Meiser2009,Norcia2016,Bohnet2012,Debnath2018,Dong2023},
the linewidth of the bad-cavity laser is dramatically narrowed as
a result of the strong atom-atom correlation. Hence the linewidth
and the central frequency of the laser spectrum can be approximated
utilizing one of the eigenvalues of $\mathbb{W}$ (the one with the
smaller real part)
\begin{align}
\Delta\nu & \simeq\frac{\kappa\Gamma-N\Omega_{d}^{2}\left\langle \hat{\sigma}_{z}\right\rangle _{\mathrm{s}}}{\kappa+\Gamma}\simeq\frac{\kappa\Omega_{d}^{2}[1-(d_{0}-1)C+C^{2}]}{(\Gamma+\kappa)^{2}C(d_{0}C-1)},\label{eq:line ana}\\
\nu & \simeq\frac{\kappa\omega_{d}+\Gamma\omega_{c}}{\kappa+\Gamma}.\label{eq:fre ana}
\end{align}
It is worth to emphasize that the analytical expression of the linewidth
is derived from the first-order correction of $\left\langle \hat{\sigma}_{z}\right\rangle _{\mathrm{s}}$
(see Appendix \ref{sec:quan regression theorem}).

\section{Output Laser Properties}

In the above section, we have examined the dynamics of the dressed-dark-state
superradiant laser system analytically which characterizes the properties
of the laser (steady-state radiation, laser spectrum, and so on).
As shown in Eqs. (\ref{eq:appro solu-1})-(\ref{eq:appro solu-4})
and (\ref{eq:line ana})-(\ref{eq:fre ana}), since the lasing state
in our scheme, i.e., the dressed-dark state, originates from the M1
transition, the behaviour of the output laser would be manipulated
by this magnetic field. As an representative example, in the following
we explore our proposal utilizing an ensemble of $^{88}$Sr atoms
to demonstrate the features of this magnetic-field-induced superradiant
laser more intuitively. Here the energy gap between the bare bright
and dark states is $\Delta=2\pi\times5.6$THz. The decay rate of the
$^{3}$P$_{1}$-$^{1}$S$_{0}$ transition is $\gamma=2\pi\times7.5$kHz.
As achieved in experiments \citep{Bohnet2012,Norcia2016,Norcia2016a},
the atomic number in the cavity can be up to $N=10^{6}$ with the
cavity dissipation rate $\kappa=2\pi\times150$kHz. The Rabi frequency
of the coupling between the $^{3}$P$_{1}$-$^{1}$S$_{0}$ transition
and the cavity mode is $\Omega_{c}=2\pi\times21.2$kHz.

\subsection{Steady-state radiation}

In the superradiant laser system, the atom-atom correlation $\left\langle \hat{\sigma}_{1}^{\dagger}\hat{\sigma}_{2}\right\rangle _{\mathrm{s}}$
is remarkable in portraying the features of the steady-state radiation.
In the lasing process, together with the atomic collective decay resulting
from the adiabatic elimination of the cavity mode, the non-collective
pump and decay of the atoms drive the atoms to some collective states
$\left|J,M\right\rangle $ (Dicke state) \citep{Dicke1954} where
$J$ and $M$ are the quantum numbers satisfying $(\hat{J}_{x}^{2}+\hat{J}_{y}^{2}+\hat{J}_{z}^{2})\left|J,M\right\rangle =J(J+1)\left|J,M\right\rangle $
and $\hat{J}_{z}\left|J,M\right\rangle =M\left|J,M\right\rangle $.
As the total radiation rate of the system in the steady state is related
to the collective correlation $\left\langle \hat{J}_{+}\hat{J}_{-}\right\rangle _{\mathrm{s}}\simeq N^{2}\left\langle \hat{\sigma}_{1}^{\dagger}\hat{\sigma}_{2}\right\rangle _{\mathrm{s}}$
\citep{Meiser2010}, the steady-state photon number inside the cavity
is proportional to the atomic correlation $\left\langle \hat{c}^{\dagger}\hat{c}\right\rangle _{\mathrm{s}}\propto N^{2}\left\langle \hat{\sigma}_{1}^{\dagger}\hat{\sigma}_{2}\right\rangle _{\mathrm{s}}$.
Subsequently, according to the input-output theory, the emission power
of the photons through the cavity mirror equals the cavity dissipation
rate times the photon energy, i.e., $P=\kappa\hbar\omega\left\langle \hat{c}^{\dagger}\hat{c}\right\rangle _{\mathrm{s}}$
(see Eqs. (\ref{eq:appro solu-1}), (\ref{eq:appro solu-3}) and Fig.
\ref{fig:2}) \citep{2008a}.

To be more specific, as demonstrated in the approximate analytical
results Eq. (\ref{eq:appro solu-1}), for a small pump rate, the generalized
collective cooperative parameter $C$ is much larger than unity. Therefore,
when the pump rate exceeds the effective decay rate of the dressed-dark
state $\gamma_{d}\equiv\gamma\Omega_{B}^{2}/\Delta^{2}$, the atom-atom
correlation $\left\langle \hat{\sigma}_{1}^{\dagger}\hat{\sigma}_{2}\right\rangle _{\mathrm{s}}$
grows from negative to positive, which marks $\gamma_{d}$ as the
lower threshold of the laser. Above this threshold, due to the pump-induced
dephasing, a competition between the establishment and destruction
of the macroscopic spin appears, and thus the photon number first
increases and then decreases with the growth of the pump. Ultimately,
the radiation ceases at the upper threshold $N\Gamma_{d,c}\equiv N\Omega_{b,0}^{2}\Omega_{B}^{2}/(\kappa\Delta^{2})$
where $\Gamma_{d,c}$ denotes the collective decay rate of the dressed-dark
state. Notably, beyond this upper threshold, the atomic population
is inverted, and the correlation between atoms vanishes \citep{Meiser2010}.

Figures \ref{fig:2}(a) and \ref{fig:2}(b) display the steady-state
radiation versus the pump rate for the cases $\Omega_{B}/\Delta=10^{-1}$
and $\Omega_{B}/\Delta=10^{-3}$, respectively. It is worth noting
that for the system we discussed here, the boundary between the aforementioned
crossover and superradiant lasing regimes is $\Omega_{B}/\Delta\simeq10^{-2.15}$
(see Tab. \ref{tab:1}). That is, $\Omega_{B}/\Delta=10^{-1}$ ($\Omega_{B}/\Delta=10^{-3}$)
corresponds to the crossover (superradiant lasing) regime. In Fig.
\ref{fig:2}, the red solid lines represent the approximate analytical
result of the two-level dressed-dark-state model (Eq. (\ref{eq:appro solu-1})),
while the triangles stand for the numerical calculation results of
the original three-level model (see Appendix \ref{sec:Three level equation}).
The gray dashed vertical lines denote two thresholds of the laser.
The coincidence of these results elucidates the validity of the two-level
dressed-dark-state model in our scheme, which can be attributed to
the negligible population of the dressed-bright state (see the black
dashed lines in Figs. \ref{fig:2}(c) and \ref{fig:2}(d)).

\begin{figure}[t]
\begin{centering}
\includegraphics[scale=0.3]{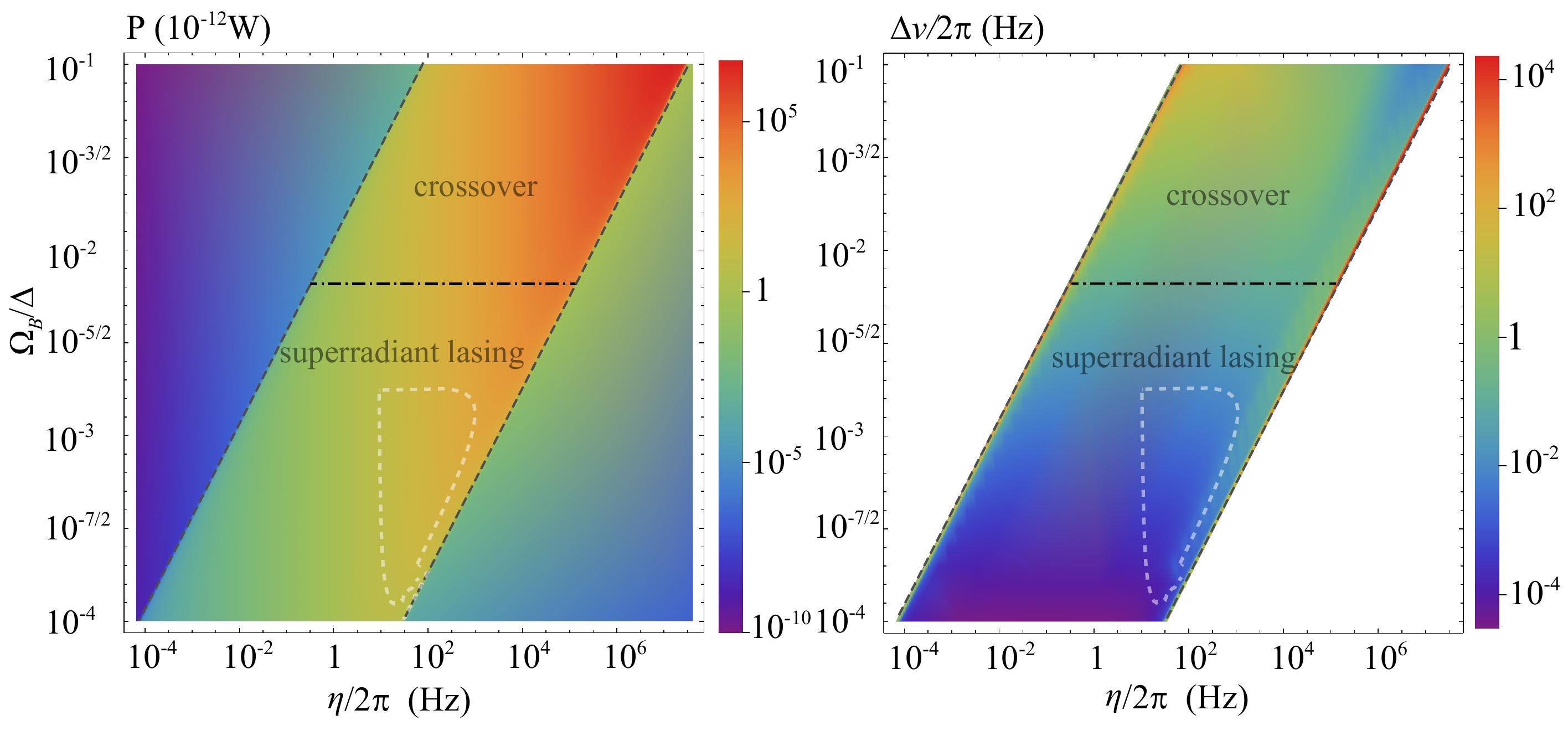}
\par\end{centering}
\caption{\label{fig:4}The contours of the power $P$ (a) and the linewidth
$\Delta\nu$ (b) versus the pump rate $\eta$ and the magnetic field
$\Omega_{B}/\Delta$. Two gray dashed lines ($\gamma_{d}\equiv\gamma\Omega_{B}^{2}/\Delta^{2}$
and $N\Gamma_{d,c}\equiv N\Omega_{b,0}^{2}\Omega_{B}^{2}/(\kappa\Delta^{2})$)
represent the thresholds of the laser system. The horizontal black
dot-dashed line ($\Omega_{B}/\Delta\simeq10^{-2.15}$) separates the
crossover and the superradiant lasing regime. The area surrounded
by the white dashed line depicts the region region satisfying $P\protect\geq10^{-11}\mathrm{W}$,
$\Delta\nu\protect\leq2\pi\times10\mathrm{mHz}$, and $\eta\protect\leq2\pi\times10^{3}\mathrm{Hz}$.}
\end{figure}

\subsection{Laser linewidth}

The approximate analytical result (Eq. (\ref{eq:line ana})) depicts
the behaviour of the laser linewidth in details. Especially, two feature
emerge as the rise of the pump: (1) a constant linewidth $\Delta\nu\simeq\Omega_{d}^{2}/\kappa$
for a comparatively small pump ($\gamma_{d}<\eta\ll\kappa,N\Gamma_{d,c}$),
which has been demonstrated in the superradiant laser system embedded
with $^{87}$Sr atoms \citep{Meiser2009}; (2) linewidth narrowed
with the increase of the pump $\Delta\nu\simeq\Omega_{d}^{2}/\eta{}^{2}$
($\kappa\ll\eta<N\Gamma_{d,c}$), resembling a Schawlow--Townes-like
behaviour where the phase diffusion diminishes gradually with the
growth of the power \citep{Haken1983}. It is worth highlighting that
the performance of the linewidth in an individual case would vary
with the magnetic field. For instance, for a relatively large magnetic
field $\Omega_{B}/\Delta=10^{-1}$ (the crossover regime) where the
thresholds of the system satisfy $\gamma_{d}=2\pi\times75\mathrm{Hz}\ll\kappa\ll N\Gamma_{d,c}=2\pi\times30\mathrm{MHz}$,
the linewidth remains several tens of kHz for a small pump and exhibits
a Schawlow--Townes-like feature when the pump goes beyond the cavity
dissipation rate (see Fig. \ref{fig:3}(a)). In contrast, a reduced
magnetic field $\Omega_{B}/\Delta=10^{-3}$ (the superradiant lasing
regime) will shift the laser working range to a lower pump zone ($\gamma_{d}=2\pi\times7.5\mathrm{mHz}\ll N\Gamma_{d,c}=2\pi\times3\mathrm{kHz}\ll\kappa$),
resulting in a constant linewidth at the mHz level throughout the
operation region (see Fig. \ref{fig:3}(b)). In Fig. \ref{fig:4},
we also illustrate the contour plot of the laser power and linewidth
versus the pump rate and magnetic field, which unveils the laser properties
more clearly.

The features of the linewidth are closely related to the coherence
and the working regimes of the system. With the coherence contribution
parameter defined in Eq. (\ref{eq:coh-contri-para}), the laser linewidth
can be recast as $\Delta\nu\simeq\Omega_{d}^{2}/[\kappa(1+\lambda)^{2}]$.
When the main noise of the system comes from the cavity ($\kappa\gg\eta$,
$\lambda\ll1$) and the cavity mode can be eliminated adiabatically,
the phase information of the system is carried by the atoms \citep{Bohnet2012},
leaving the laser linewidth a constant $\Delta\nu\sim\Omega_{d}^{2}/\kappa$
(coherence in atoms). Meanwhile, for the case of a large pump ($\eta\gg\kappa$,
$\lambda\gg1$), the intracavity photon number grows drastically,
which in turn boosts the stimulated emission process and inhibits
the phase diffusion of the light field. Consequently, the laser is
narrowed with the rise of the pump $\Delta\nu\sim\Omega_{d}^{2}/\eta{}^{2}$
(coherence mainly in photons). Therefore, the linewidth exhibits the
constant and Schawlow--Townes-like features in the crossover regime
(the coherence contribution parameter from $\lambda\ll1$ to $\lambda\gg1$)
in turn, while remaining constant in the superradiant lasing regime
($\lambda\ll1$). What is more, as a result of the shrinkage of laser
thresholds, the laser system can operate from the crossover regime
to the superradiant lasing regime with the reduction of the magnetic
field, where the laser linewidth is remarkably reduced (see the horizontal
black dot-dashed line in Fig. \ref{fig:4}).

\begin{figure}[t]
\begin{centering}
\includegraphics[scale=0.3]{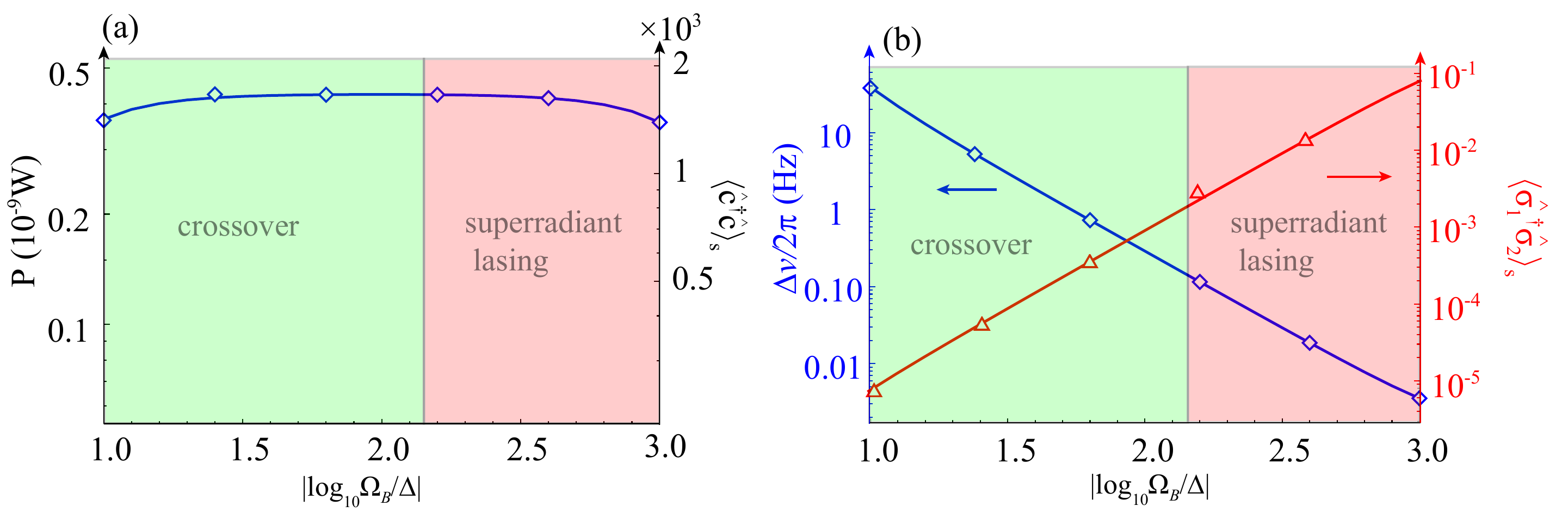}
\par\end{centering}
\caption{\label{fig:5}(a) The power maintaining and (b) linewidth narrowing
versus the magnetic field strength $\Omega_{B}/\Delta$. The solid
lines represent the approximate analytical results of the two-level
model while the triangles and diamonds stand for the numerical results
of the three-level model for $\eta=2\pi\times0.5\mathrm{kHz}$.}
\end{figure}

This line-narrowing phenomenon is more evident in Fig. \ref{fig:5}.
For a fixed pump $\eta=2\pi\times0.5\mathrm{kHz}$, the laser is continuously
narrowed with the decrease of $\Omega_{B}/\Delta$ while maintaining
its power. In this sense, one can achieve a laser with a fairly small
linewidth and large power at a small pump by operating the system
in the superradiant lasing regime. Emerging from the growth of the
atom-atom correlation (see Fig. \ref{fig:5}(b)), this ultranarrow
laser at a quite small pump mitigates the heating effect, and thus
would be beneficial to the practical use \citep{Dong2023}. For clarity,
we have marked the lasing region satisfying $P\geq10^{-11}\mathrm{W}$,
$\Delta\nu\leq2\pi\times10\mathrm{mHz}$, and $\eta\leq2\pi\times10^{3}\mathrm{Hz}$
in Figs. \ref{fig:2}-\ref{fig:4} (see the purple marked region in
Figs. \ref{fig:2} and \ref{fig:3} and the area surrounded by the
white dashed line in Fig. \ref{fig:4}). As is demonstrated, the laser
satisfying the above conditions operates in the superradiant lasing
regime.

\subsection{Frequency and pulling coefficient}

As exhibited in Sec. \ref{subsec:Laser-spectrum}, the laser frequency
refers to the central frequency of the output spectrum and is a weighted
expectation of the cavity and lasing state frequency, i.e., $\nu\simeq(\kappa\omega_{d}+\Gamma\omega_{c})/(\kappa+\Gamma)$.
Apparently, the resonance of the cavity mode with respect to the lasing
state indicates a laser frequency as that of the state (see Appendix
\ref{sec:quan regression theorem}). However, due to the inevitable
thermal fluctuation of the cavity mirror, the laser frequency may
deviate from this resonance frequency. Meanwhile, as the lasing state
in our scheme is not a bare atomic state but a dressed state whose
frequency varies with the magnetic field ($\omega_{d}=\omega_{d,0}-\varpi_{B}$
with $\varpi_{B}\equiv-\Delta/2+\sqrt{(\Delta/2)^{2}+\Omega_{B}^{2}}$),
the fluctuation of the magnetic field strength (or equivalently, the
magnetic Rabi frequency) may also pull the laser frequency away. Here
we investigate this frequency pulling effect induced by the cavity
or magnetic field fluctuations and try to uncover the frequency robustness
of the photons in the superradiant lasing regime. The pulling coefficient
can be defined as 
\begin{equation}
c_{P}^{(i)}\equiv\left|\frac{\partial(\nu^{*}-\nu)}{\partial\delta_{i}}\right|,i=c,B,
\end{equation}
where $\nu^{*}-\nu$ represents the laser frequency change as a result
of the cavity frequency shift $\delta_{c}=\omega_{c}-\omega_{d}$
or the magnetic Rabi frequency shift $\delta_{B}=\Omega_{B}^{'}-\Omega_{B}$.

\begin{figure}[t]
\begin{centering}
\includegraphics[scale=0.3]{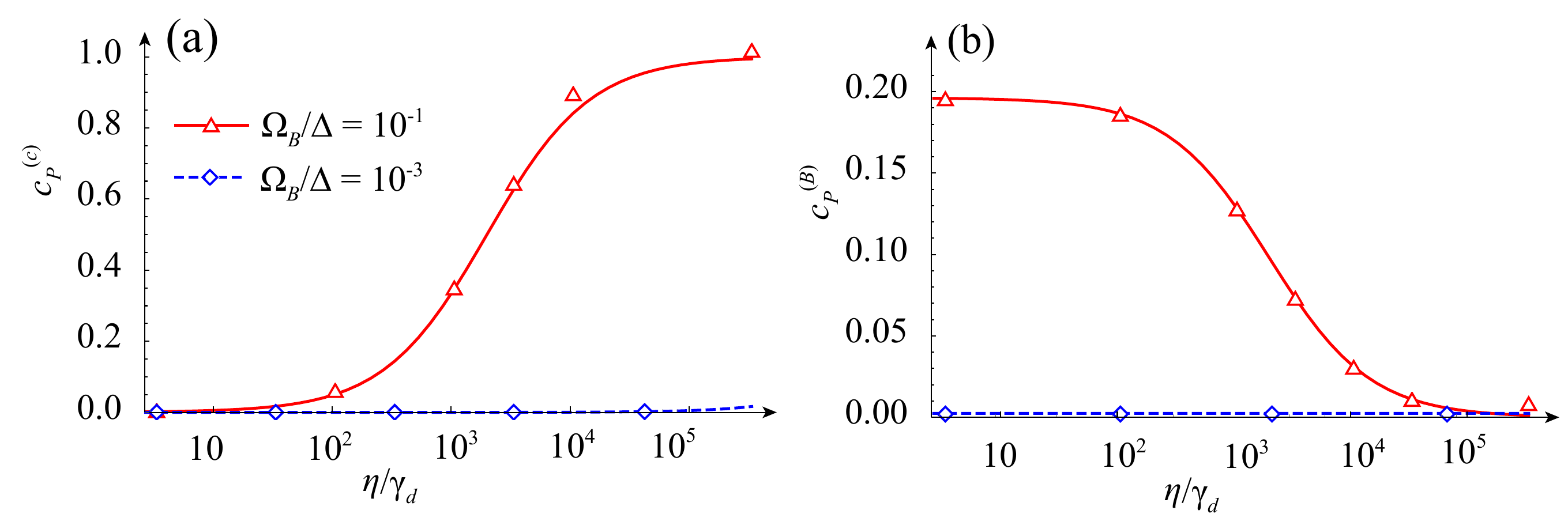}
\par\end{centering}
\caption{\label{fig:6}The cavity (a) and magnetic-field (b) pulling coefficients
versus the rescaled pump. The red solid (blue dashed) lines represent
the approximate analytical results of the two-level model, while the
triangles (diamonds) stand for the numerical results of the three-level
model for $\Omega_{B}/\Delta=10^{-1}$ ($\Omega_{B}/\Delta=10^{-3}$).
In contrast to the fairly large frequency shift in the crossover regime
($\Omega_{B}/\Delta=10^{-1}$), the laser frequency is more robust
in the superradiant lasing regime ($\Omega_{B}/\Delta=10^{-3}$).}
\end{figure}

The laser frequency pulling via the cavity and magnetic field can
be well controlled in the superradiant lasing regime. Notably, as
illustrated in Eq. (\ref{eq:fre ana}), the cavity and magnetic-field
pulling coefficients read
\begin{align}
c_{P}^{(c)} & \simeq\Gamma/(\kappa+\Gamma),\\
c_{P}^{(B)} & \simeq2\frac{\frac{\omega_{c}}{\Delta}\gamma-\kappa}{\kappa+\Gamma}\frac{\Omega_{B}}{\Delta}.\label{eq:magnetic_pull}
\end{align}
 In Fig. \ref{fig:6}, we plot the two pulling coefficients with respect
to the pump for different magnetic field cases. On one hand, owing
to the shrinkage of the laser thresholds, $c_{P}^{(c)}$ is suppressed
when the system tends to the superradiant lasing regime, i.e., $c_{P}^{(c)}\ll1$
for $\Gamma\ll\kappa$ (see Fig. \ref{fig:6}(a)). Particularly, for
$\Omega_{B}/\Delta=10^{-1}$ (the crossover regime), $c_{P}^{(c)}$
grows from $10^{-3}$ to almost unity with the rise of the pump. As
a contrast, it is no more than $0.02$ when the magnetic field is
reduced to $\Omega_{B}/\Delta=10^{-3}$ (the superradiant lasing regime).
On the other hand, the magnetic field-pulling coefficient $c_{P}^{(B)}$
is approximately proportional to the magnetic field, indicating a
relatively small pulling for a rather weak magnetic field. In Fig.
\ref{fig:6}(b), it has been demonstrated that the blue dashed line
($\Omega_{B}/\Delta=10^{-3}$, the superradiant lasing regime) is
much lower than red dashed line ($\Omega_{B}/\Delta=10^{-1}$, the
crossover regime). As a consequence, compared to the case in the crossover
regime, the laser frequency is more robust against both the cavity
and magnetic field fluctuations in the superradiant lasing regime,
which relaxes the cavity-length and field control requirement of the
scheme and brings convenience for the implementation of this scheme.

\section{Discussions and Conclusions}

In this work, we explore a magnetic-field-induced superradiant laser
scheme where the lasing state is an atomic dark state dressed with
a small bright component via the magnetic dipole transition. Hence
the laser properties, for instance, the laser thresholds, power, and
linewidth, could be manipulated by this external field. Specifically,
with the variation of the magnetic field, the laser system can operate
from the crossover (coherence in both atoms and photons) to the superradiant
lasing regimes (coherence only in atoms). In contrast to the crossover
regime, the laser in the superradiant lasing regime exhibits a significant
line-narrowing feature (mHz level) while maintaining its power. Further
research reveals that this effect emerges from the strong atom-atom
correlation generated from the atomic macroscopic spin. Moreover,
although the laser frequency would be shifted by the cavity length
and magnetic field fluctuations, it has been demonstrated that both
pulling coefficients can be well controlled as long as the system
operates in the superradiant lasing regime. This flexibility relaxes
the field and cavity-length control requirements for the experimental
implementation.

It is interesting to compare the proposal here with our previous coherent-assisted
superradiant laser scheme \citep{Dong2023}. In our previous work,
stemming from the coherence between the dressed bright and dark states
constructed by two Raman beams, a new local minimum of the linewidth
emerges at the cost of the laser power, manifesting the possibility
of realizing a laser with a comparatively small linewidth at a low
pump. However, since the laser frequency is hypersensitive to the
fluctuation of the frequency difference of the Raman beams, the scenario
in \citep{Dong2023} demands a judicious control of the Raman beams.
In our present model, the lasing state is a dressed-dark state induced
by a static magnetic field (the dressed-bright state is negligible
due to the large detuning). The laser linewidth in the present work
is reduced by the growth of the atom-atom correlation without the
loss of its power. Meanwhile, the laser frequency pulling is dramatically
mitigated. Therefore, one can achieve a laser with a rather small
linewidth and large power by operating the system deep into the superradiant
lasing regime, where the laser frequency is robust against both the
fluctuations of the cavity length and magnetic field strength.

Emerging from the magnetic-field-induced dressed-dark state, our proposal
can be performed with many of the alkaline-earth-metal-like atoms
(e.g., Sr, Mg, Ca, and Yb) and may find extensive applications in
frequency reference and precise metrology scenarios such as gravitational
wave detection and measurements of fundamental constants \citep{Abramovici1992,Harry2006,Graham2013,Fortier2007,Rosi2014}.
\begin{acknowledgments}
We thank An-An Yao for helpful discussions. G.D is supported by National
Natural Science Foundation of China (NSFC) Grant No. 12205211. Y.Y.
is supported by NSFC Grant No. 12175204.
\end{acknowledgments}

\appendix
%dummy comment inserted by tex2lyx to ensure that this paragraph is not empty%dummy comment inserted by tex2lyx to ensure that this paragraph is not empty%dummy comment inserted by tex2lyx to ensure that this paragraph is not empty%dummy comment inserted by tex2lyx to ensure that this paragraph is not empty%dummy comment inserted by tex2lyx to ensure that this paragraph is not empty
%dummy comment inserted by tex2lyx to ensure that this paragraph is not empty

\section{\label{sec:Three level equation}The Dynamical Equations for the
three-level model}

The dynamical equations for the three-level laser model shown in Fig.
\ref{fig:schematic diagram}(b) are \begin{widetext}
\begin{align}
\frac{d}{dt}\left\langle \hat{c}^{\dagger}\hat{c}\right\rangle  & =-\kappa\left\langle \hat{c}^{\dagger}\hat{c}\right\rangle +\frac{N\Omega_{d}}{2i}\left(\left\langle \hat{c}^{\dagger}\hat{\sigma}_{gb_{0}}\right\rangle -c.c.\right),\\
\frac{d}{dt}\left\langle \hat{c}^{\dagger}\hat{\sigma}_{gb_{0}}\right\rangle  & =\left[i(\delta-\Delta)-\frac{\eta+\gamma+\kappa}{2}\right]\left\langle \hat{c}^{\dagger}\hat{\sigma}_{gb_{0}}\right\rangle +i\frac{\Omega_{b,0}}{2}\left(\left\langle \hat{\sigma}_{b_{0}b_{0}}\right\rangle -\left\langle \hat{\sigma}_{gg}\right\rangle \right)\left\langle \hat{c}^{\dagger}\hat{c}\right\rangle -i\Omega_{B}\left\langle \hat{c}^{\dagger}\hat{\sigma}_{gd_{0}}\right\rangle \nonumber \\
 & +i\frac{\Omega_{d}}{2}\left\langle \hat{\sigma}_{b_{0}b_{0}}\right\rangle +i\frac{\Omega_{d}}{2}(N-1)\left\langle \hat{\sigma}_{b_{0}g}^{1}\hat{\sigma}_{gb_{0}}^{2}\right\rangle ,\\
\frac{d}{dt}\left\langle \hat{c}^{\dagger}\hat{\sigma}_{gd_{0}}\right\rangle  & =\left[i\delta-\frac{\eta+\kappa}{2}\right]\left\langle \hat{c}^{\dagger}\hat{\sigma}_{gd_{0}}\right\rangle +i\frac{\Omega_{b,0}}{2}\left\langle \hat{\sigma}_{b_{0}d_{0}}\right\rangle \left\langle \hat{c}^{\dagger}\hat{c}\right\rangle -i\Omega_{B}\left\langle \hat{c}^{\dagger}\hat{\sigma}_{gb_{0}}\right\rangle \nonumber \\
 & +i\frac{\Omega_{d}}{2}\left\langle \hat{\sigma}_{b_{0}d_{0}}\right\rangle +i\frac{\Omega_{d}}{2}(N-1)\left\langle \hat{\sigma}_{b_{0}g}^{1}\hat{\sigma}_{gd_{0}}^{2}\right\rangle ,\\
\frac{d}{dt}\left\langle \hat{\sigma}_{d_{0}b_{0}}\right\rangle  & =\left[-i\Delta-\frac{\gamma}{2}\right]\left\langle \hat{\sigma}_{d_{0}b_{0}}\right\rangle -i\frac{\Omega_{b,0}}{2}\left\langle \hat{\sigma}_{d_{0}g}\hat{c}\right\rangle -i\Omega_{B}\left(\left\langle \hat{\sigma}_{d_{0}d_{0}}\right\rangle -\left\langle \hat{\sigma}_{b_{0}b_{0}}\right\rangle \right),\\
\frac{d}{dt}\left\langle \hat{\sigma}_{b_{0}b_{0}}\right\rangle  & =-\gamma\left\langle \hat{\sigma}_{b_{0}b_{0}}\right\rangle +i\frac{\Omega_{b,0}}{2}\left(\left\langle \hat{c}^{\dagger}\hat{\sigma}_{gb_{0}}\right\rangle -c.c.\right)+i\Omega_{B}\left(\left\langle \hat{\sigma}_{d_{0}b_{0}}\right\rangle -c.c.\right),\\
\frac{d}{dt}\left\langle \hat{\sigma}_{d_{0}d_{0}}\right\rangle  & =\eta\left\langle \hat{\sigma}_{gg}\right\rangle -i\Omega_{B}\left(\left\langle \hat{\sigma}_{d_{0}b_{0}}\right\rangle -c.c.\right),\\
\frac{d}{dt}\left\langle \hat{\sigma}_{b_{0}g}^{1}\hat{\sigma}_{gb_{0}}^{2}\right\rangle  & =-(\gamma+\eta)\left\langle \hat{\sigma}_{b_{0}g}^{1}\hat{\sigma}_{gb_{0}}^{2}\right\rangle -i\frac{\Omega_{b,0}}{2}\left(\left\langle \hat{\sigma}_{b_{0}b_{0}}\right\rangle -\left\langle \hat{\sigma}_{gg}\right\rangle \right)\left(\left\langle \hat{c}^{\dagger}\hat{\sigma}_{gb_{0}}\right\rangle -c.c.\right)+i\Omega_{B}\left(\left\langle \hat{\sigma}_{d_{0}g}^{1}\hat{\sigma}_{gb_{0}}^{2}\right\rangle -c.c.\right),\\
\frac{d}{dt}\left\langle \hat{\sigma}_{d_{0}g}^{1}\hat{\sigma}_{gd_{0}}^{2}\right\rangle  & =-\eta\left\langle \hat{\sigma}_{d_{0}g}^{1}\hat{\sigma}_{gd_{0}}^{2}\right\rangle -i\frac{\Omega_{b,0}}{2}\left(\left\langle \hat{\sigma}_{d_{0}b_{0}}\right\rangle \left\langle \hat{c}^{\dagger}\hat{\sigma}_{gd_{0}}\right\rangle -c.c.\right)+i\Omega_{B}\left(\left\langle \hat{\sigma}_{b_{0}g}^{1}\hat{\sigma}_{gd_{0}}^{2}\right\rangle -c.c.\right),\\
\frac{d}{dt}\left\langle \hat{\sigma}_{b_{0}g}^{1}\hat{\sigma}_{gd_{0}}^{2}\right\rangle  & =\left[i\Delta-\frac{2\eta+\gamma}{2}\right]\left\langle \hat{\sigma}_{b_{0}g}^{1}\hat{\sigma}_{gd_{0}}^{2}\right\rangle -i\frac{\Omega_{b,0}}{2}\left(\left\langle \hat{\sigma}_{b_{0}b_{0}}\right\rangle -\left\langle \hat{\sigma}_{gg}\right\rangle \right)\left\langle \hat{c}^{\dagger}\hat{\sigma}_{gd_{0}}\right\rangle \nonumber \\
 & +i\frac{\Omega_{b,0}}{2}\left\langle \hat{\sigma}_{b_{0}d_{0}}\right\rangle \left\langle \hat{\sigma}_{b_{0}g}\hat{c}\right\rangle +i\Omega_{B}\left\langle \hat{\sigma}_{d_{0}g}^{1}\hat{\sigma}_{gd_{0}}^{2}\right\rangle -i\Omega_{B}\left\langle \hat{\sigma}_{b_{0}g}^{1}\hat{\sigma}_{gb_{0}}^{2}\right\rangle ,
\end{align}
 \end{widetext}where c.c. stands for the conjugate complex. We have
defined the detunings $\delta\equiv\omega_{c}-\omega_{d,0}$, $\Delta\equiv\omega_{b,0}-\omega_{d,0}$
and the operator $\hat{\sigma}_{nm}\equiv\left|n\right\rangle \left\langle m\right|$
for $n,m=g,b_{0},d_{0}$.

\section{\label{sec:appro analy solu}The approximate analytical solutions
of the two-level dynamical equations}

When taking the quantity $i\Omega_{d}[(\left\langle \hat{\sigma}_{z}\right\rangle +1)/2-\left\langle \hat{\sigma}_{1}^{\dagger}\hat{\sigma}_{2}\right\rangle ]/2$
into account, we obtain the high order corrections for the solutions
of Eqs. (\ref{eq:photon number})-(\ref{eq:atom-field corr}). For
example, the dynamical equations for the first-order correction can
be cast as \begin{widetext}

\begin{align}
\frac{d}{dt}\left\langle \hat{c}^{\dagger}\hat{c}\right\rangle ^{(1)} & =-\kappa\left\langle \hat{c}^{\dagger}\hat{c}\right\rangle ^{(1)}+\frac{N\Omega_{d}}{2i}\left(\left\langle \hat{c}^{\dagger}\hat{\sigma}\right\rangle ^{(1)}-\left\langle \hat{\sigma}^{\dagger}\hat{c}\right\rangle ^{(1)}\right),\\
\frac{d}{dt}\left\langle \hat{\sigma}_{z}\right\rangle ^{(1)} & =-\Gamma\left\langle \hat{\sigma}_{z}\right\rangle ^{(1)}+i\Omega_{d}\left(\left\langle \hat{c}^{\dagger}\hat{\sigma}\right\rangle ^{(1)}-\left\langle \hat{\sigma}^{\dagger}\hat{c}\right\rangle ^{(1)}\right),\\
\frac{d}{dt}\left\langle \hat{\sigma}_{1}^{\dagger}\hat{\sigma}_{2}\right\rangle ^{(1)} & =-\Gamma\left\langle \hat{\sigma}_{1}^{\dagger}\hat{\sigma}_{2}\right\rangle ^{(1)}+\frac{\Omega_{d}\left\langle \hat{\sigma}_{z}\right\rangle _{\mathrm{s}}^{(0)}}{2i}\left(\left\langle \hat{c}^{\dagger}\hat{\sigma}\right\rangle ^{(1)}-\left\langle \hat{\sigma}^{\dagger}\hat{c}\right\rangle ^{(1)}\right)+\frac{\Omega_{d}\left\langle \hat{\sigma}_{z}\right\rangle ^{(1)}}{2i}\left(\left\langle \hat{c}^{\dagger}\hat{\sigma}\right\rangle _{\mathrm{s}}^{(0)}-\left\langle \hat{\sigma}^{\dagger}\hat{c}\right\rangle _{\mathrm{s}}^{(0)}\right),\\
\frac{d}{dt}\left\langle \hat{c}^{\dagger}\hat{\sigma}\right\rangle ^{(1)} & =-\left(\frac{\Gamma+\kappa}{2}\right)\left\langle \hat{c}^{\dagger}\hat{\sigma}\right\rangle ^{(1)}+i\frac{\Omega_{d}}{2}\left(\left\langle \hat{c}^{\dagger}\hat{c}\right\rangle _{\mathrm{s}}^{(0)}\left\langle \hat{\sigma}_{z}\right\rangle ^{(1)}+\left\langle \hat{\sigma}_{z}\right\rangle _{\mathrm{s}}^{(0)}\left\langle \hat{c}^{\dagger}\hat{c}\right\rangle ^{(1)}+N\left\langle \hat{\sigma}_{1}^{\dagger}\hat{\sigma}_{2}\right\rangle ^{(1)}\right)\nonumber \\
 & +i\frac{\Omega_{d}}{2}\left(\frac{\left\langle \hat{\sigma}_{z}\right\rangle _{\mathrm{s}}^{(0)}+1}{2}-\left\langle \hat{\sigma}_{1}^{\dagger}\hat{\sigma}_{2}\right\rangle _{\mathrm{s}}^{(0)}\right).
\end{align}

\end{widetext} with the steady-state solution
\begin{align}
\left\langle \hat{c}^{\dagger}\hat{c}\right\rangle _{\mathrm{s}}^{(1)} & =\frac{\Gamma[1-(d_{0}-1)C+C^{2}]}{2(\Gamma+\kappa)C(d_{0}-1)},\\
\left\langle \hat{\sigma}_{z}\right\rangle _{\mathrm{s}}^{(1)} & =-\frac{\kappa[1-(d_{0}-1)C+C^{2}]}{N(\Gamma+\kappa)C(d_{0}C-1)},\\
\left\langle \hat{\sigma}_{1}^{\dagger}\hat{\sigma}_{2}\right\rangle _{\mathrm{s}}^{(1)} & =\frac{\kappa(2-d_{0}C)[1-(d_{0}-1)C+C^{2}]}{2N(\Gamma+\kappa)C^{2}(d_{0}C-1)},\\
\left\langle \hat{c}^{\dagger}\hat{\sigma}\right\rangle _{\mathrm{s}}^{(1)} & =i\frac{[1-(d_{0}-1)C+C^{2}]\Omega_{d}}{2(\Gamma+\kappa)C^{2}(d_{0}C-1)}.
\end{align}

\section{quantum regression theorem\label{sec:quan regression theorem}}

As we have shown in Sec. \ref{subsec:Laser-spectrum}, the linewidth
and central frequency of the output laser can be characterized by
the eigenvalue of the matrix with the smaller real part. The eigen
function of the matrix $\mathbb{\mathbb{W}}$ reads \citep{Dong2023}
\begin{align}
\mathbb{\mathbb{W}}\left|j\right\rangle =w_{j}\left|j\right\rangle , & \left\langle \widetilde{j}\right|\mathbb{W}=\left\langle \widetilde{j}\right|w_{j},
\end{align}
where $w_{j}$ denotes the $j$th eigenvalue of the matrix with the
right (left) eigenvector $\left|j\right\rangle $ ($\left\langle \widetilde{j}\right|$).
Moreover, the eigenvectors of the matrix $\mathbb{W}$ are orthogonal
and normalized to each other as $\left\langle \widetilde{j}|i\right\rangle =\delta_{i,j}$
for $i,j=1,2$. The function for the eigenvalue $w_{j}$ is
\begin{equation}
\left(2w_{j}-2i\delta_{c}+\kappa\right)\left(2w_{j}+\Gamma\right)-N\Omega_{d}^{2}\left\langle \hat{\sigma}_{z}\right\rangle _{\mathrm{s}}=0.
\end{equation}
Thus the eigenvalue with the smaller real part is 
\begin{align}
w_{\mathrm{min}} & \simeq\frac{N\Omega_{d}^{2}\left\langle \hat{\sigma}_{z}\right\rangle _{\mathrm{s}}-\Gamma(\kappa-2i\delta_{c})}{2(\Gamma+\kappa-2i\delta_{c})}\simeq\frac{N\Omega_{d}^{2}\left\langle \hat{\sigma}_{z}\right\rangle _{\mathrm{s}}-\Gamma\kappa}{2(\Gamma+\kappa)}+i\frac{\Gamma\delta_{c}}{\Gamma+\kappa}.
\end{align}
In the last step, we have assumed a small detuning.

\begin{figure}[t]
\begin{centering}
\includegraphics[scale=0.3]{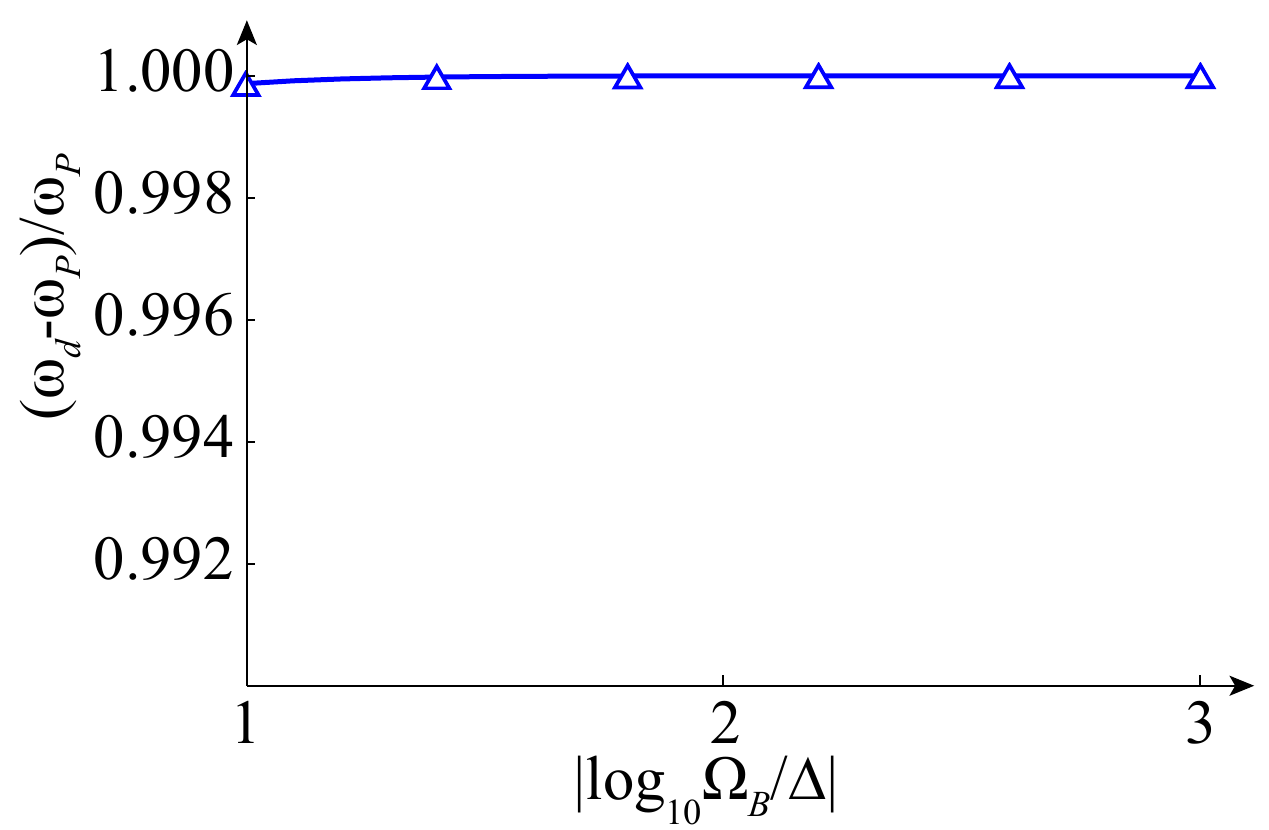}
\par\end{centering}
\caption{\label{fig:7}The laser frequency versus the magnetic field. The solid
line represents the approximated analytical results of the two-level
model while the triangles stand for the numerical results of the three-level
model. Here we choose $\eta=2\pi\times0.5$kHz.}
\end{figure}

The linewidth and central frequency of the output laser are 
\begin{align}
\Delta\nu & \simeq2\left|\mathrm{Re}(w_{\mathrm{min}})\right|\simeq\frac{\kappa\Omega_{d}^{2}[1-(d_{0}-1)C+C^{2}]}{(\Gamma+\kappa)^{2}C(d_{0}C-1)},\\
\nu & \simeq\omega_{d}+\mathrm{Im}(w_{\mathrm{min}})=\frac{\Gamma\omega_{c}+\kappa\omega_{d}}{\Gamma+\kappa},
\end{align}
which is Eq. (\ref{eq:fre ana}) in the main text. In Fig. \ref{fig:7},
we plot the frequency of the output laser which demonstrates the validity
of our approximated analytical solutions.


\begin{thebibliography}{37}%
	\makeatletter
	\providecommand \@ifxundefined [1]{%
		\@ifx{#1\undefined}
	}%
	\providecommand \@ifnum [1]{%
		\ifnum #1\expandafter \@firstoftwo
		\else \expandafter \@secondoftwo
		\fi
	}%
	\providecommand \@ifx [1]{%
		\ifx #1\expandafter \@firstoftwo
		\else \expandafter \@secondoftwo
		\fi
	}%
	\providecommand \natexlab [1]{#1}%
	\providecommand \enquote  [1]{``#1''}%
	\providecommand \bibnamefont  [1]{#1}%
	\providecommand \bibfnamefont [1]{#1}%
	\providecommand \citenamefont [1]{#1}%
	\providecommand \href@noop [0]{\@secondoftwo}%
	\providecommand \href [0]{\begingroup \@sanitize@url \@href}%
	\providecommand \@href[1]{\@@startlink{#1}\@@href}%
	\providecommand \@@href[1]{\endgroup#1\@@endlink}%
	\providecommand \@sanitize@url [0]{\catcode `\\12\catcode `\$12\catcode `\&12\catcode `\#12\catcode `\^12\catcode `\_12\catcode `\%12\relax}%
	\providecommand \@@startlink[1]{}%
	\providecommand \@@endlink[0]{}%
	\providecommand \url  [0]{\begingroup\@sanitize@url \@url }%
	\providecommand \@url [1]{\endgroup\@href {#1}{\urlprefix }}%
	\providecommand \urlprefix  [0]{URL }%
	\providecommand \Eprint [0]{\href }%
	\providecommand \doibase [0]{http://dx.doi.org/}%
	\providecommand \selectlanguage [0]{\@gobble}%
	\providecommand \bibinfo  [0]{\@secondoftwo}%
	\providecommand \bibfield  [0]{\@secondoftwo}%
	\providecommand \translation [1]{[#1]}%
	\providecommand \BibitemOpen [0]{}%
	\providecommand \bibitemStop [0]{}%
	\providecommand \bibitemNoStop [0]{.\EOS\space}%
	\providecommand \EOS [0]{\spacefactor3000\relax}%
	\providecommand \BibitemShut  [1]{\csname bibitem#1\endcsname}%
	\let\auto@bib@innerbib\@empty
	%</preamble>
	\bibitem [{\citenamefont {Maiman}(1960)}]{MAIMAN1960}%
	\BibitemOpen
	\bibfield  {author} {\bibinfo {author} {\bibfnamefont {T.~H.}\ \bibnamefont {Maiman}},\ }\href {\doibase 10.1038/187493a0} {\bibfield  {journal} {\bibinfo  {journal} {Nature}\ }\textbf {\bibinfo {volume} {187}},\ \bibinfo {pages} {493} (\bibinfo {year} {1960})}\BibitemShut {NoStop}%
	\bibitem [{\citenamefont {Derevianko}\ and\ \citenamefont {Katori}(2011)}]{Derevianko2011}%
	\BibitemOpen
	\bibfield  {author} {\bibinfo {author} {\bibfnamefont {A.}~\bibnamefont {Derevianko}}\ and\ \bibinfo {author} {\bibfnamefont {H.}~\bibnamefont {Katori}},\ }\href {\doibase 10.1103/revmodphys.83.331} {\bibfield  {journal} {\bibinfo  {journal} {Rev. Mod. Phys.}\ }\textbf {\bibinfo {volume} {83}},\ \bibinfo {pages} {331} (\bibinfo {year} {2011})}\BibitemShut {NoStop}%
	\bibitem [{\citenamefont {Ludlow}\ \emph {et~al.}(2015)\citenamefont {Ludlow}, \citenamefont {Boyd}, \citenamefont {Ye}, \citenamefont {Peik},\ and\ \citenamefont {Schmidt}}]{Ludlow2015}%
	\BibitemOpen
	\bibfield  {author} {\bibinfo {author} {\bibfnamefont {A.~D.}\ \bibnamefont {Ludlow}}, \bibinfo {author} {\bibfnamefont {M.~M.}\ \bibnamefont {Boyd}}, \bibinfo {author} {\bibfnamefont {J.}~\bibnamefont {Ye}}, \bibinfo {author} {\bibfnamefont {E.}~\bibnamefont {Peik}}, \ and\ \bibinfo {author} {\bibfnamefont {P.~O.}\ \bibnamefont {Schmidt}},\ }\href {\doibase 10.1103/revmodphys.87.637} {\bibfield  {journal} {\bibinfo  {journal} {Rev. Mod. Phys.}\ }\textbf {\bibinfo {volume} {87}},\ \bibinfo {pages} {637} (\bibinfo {year} {2015})}\BibitemShut {NoStop}%
	\bibitem [{\citenamefont {H\"ansch}\ and\ \citenamefont {Schawlow}(1975)}]{atomcooling1975}%
	\BibitemOpen
	\bibfield  {author} {\bibinfo {author} {\bibfnamefont {T.}~\bibnamefont {H\"ansch}}\ and\ \bibinfo {author} {\bibfnamefont {A.}~\bibnamefont {Schawlow}},\ }\href@noop {} {\bibfield  {journal} {\bibinfo  {journal} {Opt. Commun.}\ }\textbf {\bibinfo {volume} {13}} (\bibinfo {year} {1975})}\BibitemShut {NoStop}%
	\bibitem [{\citenamefont {Wineland}\ and\ \citenamefont {Dehmelt}(1975)}]{ioncooling1975}%
	\BibitemOpen
	\bibfield  {author} {\bibinfo {author} {\bibfnamefont {D.~J.}\ \bibnamefont {Wineland}}\ and\ \bibinfo {author} {\bibfnamefont {H.}~\bibnamefont {Dehmelt}},\ }\href@noop {} {\bibfield  {journal} {\bibinfo  {journal} {Bull. Am. Phys. Soc.}\ }\textbf {\bibinfo {volume} {20}} (\bibinfo {year} {1975})}\BibitemShut {NoStop}%
	\bibitem [{\citenamefont {Cohen-Tannoudji}\ and\ \citenamefont {Phillips}(1990)}]{Cohen_Tannoudji1990}%
	\BibitemOpen
	\bibfield  {author} {\bibinfo {author} {\bibfnamefont {C.~N.}\ \bibnamefont {Cohen-Tannoudji}}\ and\ \bibinfo {author} {\bibfnamefont {W.~D.}\ \bibnamefont {Phillips}},\ }\href@noop {} {\bibfield  {journal} {\bibinfo  {journal} {Phys. Today}\ }\textbf {\bibinfo {volume} {43}} (\bibinfo {year} {1990})}\BibitemShut {NoStop}%
	\bibitem [{\citenamefont {Numata}\ \emph {et~al.}(2004)\citenamefont {Numata}, \citenamefont {Kemery},\ and\ \citenamefont {Camp}}]{Numata2004}%
	\BibitemOpen
	\bibfield  {author} {\bibinfo {author} {\bibfnamefont {K.}~\bibnamefont {Numata}}, \bibinfo {author} {\bibfnamefont {A.}~\bibnamefont {Kemery}}, \ and\ \bibinfo {author} {\bibfnamefont {J.}~\bibnamefont {Camp}},\ }\href {\doibase 10.1103/physrevlett.93.250602} {\bibfield  {journal} {\bibinfo  {journal} {Phys. Rev. Lett.}\ }\textbf {\bibinfo {volume} {93}},\ \bibinfo {pages} {250602} (\bibinfo {year} {2004})}\BibitemShut {NoStop}%
	\bibitem [{\citenamefont {Yu}\ \emph {et~al.}(2019)\citenamefont {Yu}, \citenamefont {Qin}, \citenamefont {Yan}, \citenamefont {Lu},\ and\ \citenamefont {Jia}}]{Yu2019}%
	\BibitemOpen
	\bibfield  {author} {\bibinfo {author} {\bibfnamefont {J.}~\bibnamefont {Yu}}, \bibinfo {author} {\bibfnamefont {Y.}~\bibnamefont {Qin}}, \bibinfo {author} {\bibfnamefont {Z.}~\bibnamefont {Yan}}, \bibinfo {author} {\bibfnamefont {H.}~\bibnamefont {Lu}}, \ and\ \bibinfo {author} {\bibfnamefont {X.}~\bibnamefont {Jia}},\ }\href {\doibase 10.1364/oe.27.003247} {\bibfield  {journal} {\bibinfo  {journal} {Opt. Express}\ }\textbf {\bibinfo {volume} {27}},\ \bibinfo {pages} {3247} (\bibinfo {year} {2019})}\BibitemShut {NoStop}%
	\bibitem [{\citenamefont {Storz}\ \emph {et~al.}(1998)\citenamefont {Storz}, \citenamefont {Braxmaier}, \citenamefont {J\"ack}, \citenamefont {Pradl},\ and\ \citenamefont {Schiller}}]{Storz1998}%
	\BibitemOpen
	\bibfield  {author} {\bibinfo {author} {\bibfnamefont {R.}~\bibnamefont {Storz}}, \bibinfo {author} {\bibfnamefont {C.}~\bibnamefont {Braxmaier}}, \bibinfo {author} {\bibfnamefont {K.}~\bibnamefont {J\"ack}}, \bibinfo {author} {\bibfnamefont {O.}~\bibnamefont {Pradl}}, \ and\ \bibinfo {author} {\bibfnamefont {S.}~\bibnamefont {Schiller}},\ }\href {\doibase 10.1364/ol.23.001031} {\bibfield  {journal} {\bibinfo  {journal} {Opt. Lett.}\ }\textbf {\bibinfo {volume} {23}},\ \bibinfo {pages} {1031} (\bibinfo {year} {1998})}\BibitemShut {NoStop}%
	\bibitem [{\citenamefont {Kessler}\ \emph {et~al.}(2012)\citenamefont {Kessler}, \citenamefont {Hagemann}, \citenamefont {Grebing}, \citenamefont {Legero}, \citenamefont {Sterr}, \citenamefont {Riehle}, \citenamefont {Martin}, \citenamefont {Chen},\ and\ \citenamefont {Ye}}]{Kessler2012}%
	\BibitemOpen
	\bibfield  {author} {\bibinfo {author} {\bibfnamefont {T.}~\bibnamefont {Kessler}}, \bibinfo {author} {\bibfnamefont {C.}~\bibnamefont {Hagemann}}, \bibinfo {author} {\bibfnamefont {C.}~\bibnamefont {Grebing}}, \bibinfo {author} {\bibfnamefont {T.}~\bibnamefont {Legero}}, \bibinfo {author} {\bibfnamefont {U.}~\bibnamefont {Sterr}}, \bibinfo {author} {\bibfnamefont {F.}~\bibnamefont {Riehle}}, \bibinfo {author} {\bibfnamefont {M.~J.}\ \bibnamefont {Martin}}, \bibinfo {author} {\bibfnamefont {L.}~\bibnamefont {Chen}}, \ and\ \bibinfo {author} {\bibfnamefont {J.}~\bibnamefont {Ye}},\ }\href {\doibase 10.1038/nphoton.2012.217} {\bibfield  {journal} {\bibinfo  {journal} {Nature Photon.}\ }\textbf {\bibinfo {volume} {6}},\ \bibinfo {pages} {687} (\bibinfo {year} {2012})}\BibitemShut {NoStop}%
	\bibitem [{\citenamefont {Meiser}\ \emph {et~al.}(2009)\citenamefont {Meiser}, \citenamefont {Ye}, \citenamefont {Carlson},\ and\ \citenamefont {Holland}}]{Meiser2009}%
	\BibitemOpen
	\bibfield  {author} {\bibinfo {author} {\bibfnamefont {D.}~\bibnamefont {Meiser}}, \bibinfo {author} {\bibfnamefont {J.}~\bibnamefont {Ye}}, \bibinfo {author} {\bibfnamefont {D.~R.}\ \bibnamefont {Carlson}}, \ and\ \bibinfo {author} {\bibfnamefont {M.~J.}\ \bibnamefont {Holland}},\ }\href {\doibase 10.1103/physrevlett.102.163601} {\bibfield  {journal} {\bibinfo  {journal} {Phys. Rev. Lett.}\ }\textbf {\bibinfo {volume} {102}},\ \bibinfo {pages} {163601} (\bibinfo {year} {2009})}\BibitemShut {NoStop}%
	\bibitem [{\citenamefont {Meiser}\ and\ \citenamefont {Holland}(2010{\natexlab{a}})}]{Meiser2010}%
	\BibitemOpen
	\bibfield  {author} {\bibinfo {author} {\bibfnamefont {D.}~\bibnamefont {Meiser}}\ and\ \bibinfo {author} {\bibfnamefont {M.~J.}\ \bibnamefont {Holland}},\ }\href {\doibase 10.1103/physreva.81.033847} {\bibfield  {journal} {\bibinfo  {journal} {Phys. Rev. A}\ }\textbf {\bibinfo {volume} {81}},\ \bibinfo {pages} {033847} (\bibinfo {year} {2010}{\natexlab{a}})}\BibitemShut {NoStop}%
	\bibitem [{\citenamefont {Meiser}\ and\ \citenamefont {Holland}(2010{\natexlab{b}})}]{Meiser2010a}%
	\BibitemOpen
	\bibfield  {author} {\bibinfo {author} {\bibfnamefont {D.}~\bibnamefont {Meiser}}\ and\ \bibinfo {author} {\bibfnamefont {M.~J.}\ \bibnamefont {Holland}},\ }\href {\doibase 10.1103/physreva.81.063827} {\bibfield  {journal} {\bibinfo  {journal} {Phys. Rev. A}\ }\textbf {\bibinfo {volume} {81}},\ \bibinfo {pages} {063827} (\bibinfo {year} {2010}{\natexlab{b}})}\BibitemShut {NoStop}%
	\bibitem [{\citenamefont {Liu}\ \emph {et~al.}(2020)\citenamefont {Liu}, \citenamefont {J\"ager}, \citenamefont {Yu}, \citenamefont {Touzard}, \citenamefont {Shankar}, \citenamefont {Holland},\ and\ \citenamefont {Nicholson}}]{Liu2020}%
	\BibitemOpen
	\bibfield  {author} {\bibinfo {author} {\bibfnamefont {H.}~\bibnamefont {Liu}}, \bibinfo {author} {\bibfnamefont {S.~B.}\ \bibnamefont {J\"ager}}, \bibinfo {author} {\bibfnamefont {X.}~\bibnamefont {Yu}}, \bibinfo {author} {\bibfnamefont {S.}~\bibnamefont {Touzard}}, \bibinfo {author} {\bibfnamefont {A.}~\bibnamefont {Shankar}}, \bibinfo {author} {\bibfnamefont {M.~J.}\ \bibnamefont {Holland}}, \ and\ \bibinfo {author} {\bibfnamefont {T.~L.}\ \bibnamefont {Nicholson}},\ }\href {\doibase 10.1103/physrevlett.125.253602} {\bibfield  {journal} {\bibinfo  {journal} {Phys. Rev. Lett.}\ }\textbf {\bibinfo {volume} {125}},\ \bibinfo {pages} {253602} (\bibinfo {year} {2020})}\BibitemShut {NoStop}%
	\bibitem [{\citenamefont {Norcia}\ and\ \citenamefont {Thompson}(2016)}]{Norcia2016}%
	\BibitemOpen
	\bibfield  {author} {\bibinfo {author} {\bibfnamefont {M.~A.}\ \bibnamefont {Norcia}}\ and\ \bibinfo {author} {\bibfnamefont {J.~K.}\ \bibnamefont {Thompson}},\ }\href {\doibase 10.1103/physrevx.6.011025} {\bibfield  {journal} {\bibinfo  {journal} {Phys. Rev. X}\ }\textbf {\bibinfo {volume} {6}},\ \bibinfo {pages} {011025} (\bibinfo {year} {2016})}\BibitemShut {NoStop}%
	\bibitem [{\citenamefont {Tieri}\ \emph {et~al.}(2017)\citenamefont {Tieri}, \citenamefont {Xu}, \citenamefont {Meiser}, \citenamefont {Cooper},\ and\ \citenamefont {Holland}}]{Holland2017}%
	\BibitemOpen
	\bibfield  {author} {\bibinfo {author} {\bibfnamefont {D.~A.}\ \bibnamefont {Tieri}}, \bibinfo {author} {\bibfnamefont {M.}~\bibnamefont {Xu}}, \bibinfo {author} {\bibfnamefont {D.}~\bibnamefont {Meiser}}, \bibinfo {author} {\bibfnamefont {J.}~\bibnamefont {Cooper}}, \ and\ \bibinfo {author} {\bibfnamefont {M.~J.}\ \bibnamefont {Holland}},\ }\href@noop {} {\bibfield  {journal} {\bibinfo  {journal} {arXiv:1702.04830}\ } (\bibinfo {year} {2017})}\BibitemShut {NoStop}%
	\bibitem [{\citenamefont {Debnath}\ \emph {et~al.}(2018)\citenamefont {Debnath}, \citenamefont {Zhang},\ and\ \citenamefont {M{\o}lmer}}]{Debnath2018}%
	\BibitemOpen
	\bibfield  {author} {\bibinfo {author} {\bibfnamefont {K.}~\bibnamefont {Debnath}}, \bibinfo {author} {\bibfnamefont {Y.}~\bibnamefont {Zhang}}, \ and\ \bibinfo {author} {\bibfnamefont {K.}~\bibnamefont {M{\o}lmer}},\ }\href {\doibase 10.1103/physreva.98.063837} {\bibfield  {journal} {\bibinfo  {journal} {Phys. Rev. A}\ }\textbf {\bibinfo {volume} {98}},\ \bibinfo {pages} {063837} (\bibinfo {year} {2018})}\BibitemShut {NoStop}%
	\bibitem [{\citenamefont {Santra}\ \emph {et~al.}(2004)\citenamefont {Santra}, \citenamefont {Christ},\ and\ \citenamefont {Greene}}]{Santra2004}%
	\BibitemOpen
	\bibfield  {author} {\bibinfo {author} {\bibfnamefont {R.}~\bibnamefont {Santra}}, \bibinfo {author} {\bibfnamefont {K.~V.}\ \bibnamefont {Christ}}, \ and\ \bibinfo {author} {\bibfnamefont {C.~H.}\ \bibnamefont {Greene}},\ }\href {\doibase 10.1103/physreva.69.042510} {\bibfield  {journal} {\bibinfo  {journal} {Phys. Rev. A}\ }\textbf {\bibinfo {volume} {69}},\ \bibinfo {pages} {042510} (\bibinfo {year} {2004})}\BibitemShut {NoStop}%
	\bibitem [{\citenamefont {Santra}\ \emph {et~al.}(2005)\citenamefont {Santra}, \citenamefont {Arimondo}, \citenamefont {Ido}, \citenamefont {Greene},\ and\ \citenamefont {Ye}}]{Santra2005}%
	\BibitemOpen
	\bibfield  {author} {\bibinfo {author} {\bibfnamefont {R.}~\bibnamefont {Santra}}, \bibinfo {author} {\bibfnamefont {E.}~\bibnamefont {Arimondo}}, \bibinfo {author} {\bibfnamefont {T.}~\bibnamefont {Ido}}, \bibinfo {author} {\bibfnamefont {C.~H.}\ \bibnamefont {Greene}}, \ and\ \bibinfo {author} {\bibfnamefont {J.}~\bibnamefont {Ye}},\ }\href {\doibase 10.1103/physrevlett.94.173002} {\bibfield  {journal} {\bibinfo  {journal} {Phys. Rev. Lett.}\ }\textbf {\bibinfo {volume} {94}},\ \bibinfo {pages} {173002} (\bibinfo {year} {2005})}\BibitemShut {NoStop}%
	\bibitem [{\citenamefont {Barber}\ \emph {et~al.}(2006)\citenamefont {Barber}, \citenamefont {Hoyt}, \citenamefont {Oates}, \citenamefont {Hollberg}, \citenamefont {Taichenachev},\ and\ \citenamefont {Yudin}}]{Barber2006}%
	\BibitemOpen
	\bibfield  {author} {\bibinfo {author} {\bibfnamefont {Z.~W.}\ \bibnamefont {Barber}}, \bibinfo {author} {\bibfnamefont {C.~W.}\ \bibnamefont {Hoyt}}, \bibinfo {author} {\bibfnamefont {C.~W.}\ \bibnamefont {Oates}}, \bibinfo {author} {\bibfnamefont {L.}~\bibnamefont {Hollberg}}, \bibinfo {author} {\bibfnamefont {A.~V.}\ \bibnamefont {Taichenachev}}, \ and\ \bibinfo {author} {\bibfnamefont {V.~I.}\ \bibnamefont {Yudin}},\ }\href {\doibase 10.1103/physrevlett.96.083002} {\bibfield  {journal} {\bibinfo  {journal} {Phys. Rev. Lett.}\ }\textbf {\bibinfo {volume} {96}},\ \bibinfo {pages} {083002} (\bibinfo {year} {2006})}\BibitemShut {NoStop}%
	\bibitem [{\citenamefont {Dong}\ \emph {et~al.}(2023)\citenamefont {Dong}, \citenamefont {Yao}, \citenamefont {Zhang},\ and\ \citenamefont {Xu}}]{Dong2023}%
	\BibitemOpen
	\bibfield  {author} {\bibinfo {author} {\bibfnamefont {G.~H.}\ \bibnamefont {Dong}}, \bibinfo {author} {\bibfnamefont {Y.}~\bibnamefont {Yao}}, \bibinfo {author} {\bibfnamefont {P.}~\bibnamefont {Zhang}}, \ and\ \bibinfo {author} {\bibfnamefont {D.~Z.}\ \bibnamefont {Xu}},\ }\href {\doibase 10.1103/physreva.107.063709} {\bibfield  {journal} {\bibinfo  {journal} {Phys. Rev. A}\ }\textbf {\bibinfo {volume} {107}},\ \bibinfo {pages} {063709} (\bibinfo {year} {2023})}\BibitemShut {NoStop}%
	\bibitem [{\citenamefont {Taichenachev}\ \emph {et~al.}(2006)\citenamefont {Taichenachev}, \citenamefont {Yudin}, \citenamefont {Oates}, \citenamefont {Hoyt}, \citenamefont {Barber},\ and\ \citenamefont {Hollberg}}]{Taichenachev2006}%
	\BibitemOpen
	\bibfield  {author} {\bibinfo {author} {\bibfnamefont {A.}~\bibnamefont {Taichenachev}}, \bibinfo {author} {\bibfnamefont {V.}~\bibnamefont {Yudin}}, \bibinfo {author} {\bibfnamefont {C.}~\bibnamefont {Oates}}, \bibinfo {author} {\bibfnamefont {C.}~\bibnamefont {Hoyt}}, \bibinfo {author} {\bibfnamefont {Z.}~\bibnamefont {Barber}}, \ and\ \bibinfo {author} {\bibfnamefont {L.}~\bibnamefont {Hollberg}},\ }\href {\doibase 10.1103/physrevlett.96.083001} {\bibfield  {journal} {\bibinfo  {journal} {Phys. Rev. Lett.}\ }\textbf {\bibinfo {volume} {96}},\ \bibinfo {pages} {083001} (\bibinfo {year} {2006})}\BibitemShut {NoStop}%
	\bibitem [{\citenamefont {Dubey}\ \emph {et~al.}(2025)\citenamefont {Dubey}, \citenamefont {Kazakov}, \citenamefont {Heizenreder}, \citenamefont {Zhou}, \citenamefont {Bennetts}, \citenamefont {Sch\"affer}, \citenamefont {Sitaram},\ and\ \citenamefont {Schreck}}]{Dubey2025}%
	\BibitemOpen
	\bibfield  {author} {\bibinfo {author} {\bibfnamefont {S.}~\bibnamefont {Dubey}}, \bibinfo {author} {\bibfnamefont {G.~A.}\ \bibnamefont {Kazakov}}, \bibinfo {author} {\bibfnamefont {B.}~\bibnamefont {Heizenreder}}, \bibinfo {author} {\bibfnamefont {S.}~\bibnamefont {Zhou}}, \bibinfo {author} {\bibfnamefont {S.}~\bibnamefont {Bennetts}}, \bibinfo {author} {\bibfnamefont {S.~A.}\ \bibnamefont {Sch\"affer}}, \bibinfo {author} {\bibfnamefont {A.}~\bibnamefont {Sitaram}}, \ and\ \bibinfo {author} {\bibfnamefont {F.}~\bibnamefont {Schreck}},\ }\href {\doibase 10.1103/physrevresearch.7.013292} {\bibfield  {journal} {\bibinfo  {journal} {Phys. Rev. Research}\ }\textbf {\bibinfo {volume} {7}},\ \bibinfo {pages} {013292} (\bibinfo {year} {2025})}\BibitemShut {NoStop}%
	\bibitem [{\citenamefont {Courtillot}\ \emph {et~al.}(2005)\citenamefont {Courtillot}, \citenamefont {Quessada-Vial}, \citenamefont {Brusch}, \citenamefont {Kolker}, \citenamefont {Rovera},\ and\ \citenamefont {Lemonde}}]{Courtillot2005}%
	\BibitemOpen
	\bibfield  {author} {\bibinfo {author} {\bibfnamefont {I.}~\bibnamefont {Courtillot}}, \bibinfo {author} {\bibfnamefont {A.}~\bibnamefont {Quessada-Vial}}, \bibinfo {author} {\bibfnamefont {A.}~\bibnamefont {Brusch}}, \bibinfo {author} {\bibfnamefont {D.}~\bibnamefont {Kolker}}, \bibinfo {author} {\bibfnamefont {G.~D.}\ \bibnamefont {Rovera}}, \ and\ \bibinfo {author} {\bibfnamefont {P.}~\bibnamefont {Lemonde}},\ }\href {\doibase 10.1140/epjd/e2005-00058-0} {\bibfield  {journal} {\bibinfo  {journal} {Eur. Phys. J. D}\ }\textbf {\bibinfo {volume} {33}},\ \bibinfo {pages} {161} (\bibinfo {year} {2005})}\BibitemShut {NoStop}%
	\bibitem [{\citenamefont {Wiener}(1930)}]{Wiener1930}%
	\BibitemOpen
	\bibfield  {author} {\bibinfo {author} {\bibfnamefont {N.}~\bibnamefont {Wiener}},\ }\href {\doibase 10.1007/bf02546511} {\bibfield  {journal} {\bibinfo  {journal} {Acta Math.}\ }\textbf {\bibinfo {volume} {55}},\ \bibinfo {pages} {117} (\bibinfo {year} {1930})}\BibitemShut {NoStop}%
	\bibitem [{\citenamefont {Khintchine}(1934)}]{Khintchine1934}%
	\BibitemOpen
	\bibfield  {author} {\bibinfo {author} {\bibfnamefont {A.}~\bibnamefont {Khintchine}},\ }\href {\doibase 10.1007/bf01449156} {\bibfield  {journal} {\bibinfo  {journal} {Math. Ann.}\ }\textbf {\bibinfo {volume} {109}},\ \bibinfo {pages} {604} (\bibinfo {year} {1934})}\BibitemShut {NoStop}%
	\bibitem [{\citenamefont {Huang}(2009)}]{Huang2009}%
	\BibitemOpen
	\bibfield  {author} {\bibinfo {author} {\bibfnamefont {K.}~\bibnamefont {Huang}},\ }\href@noop {} {\emph {\bibinfo {title} {Introduction to Statistical Physics}}}\ (\bibinfo  {publisher} {Chapman and Hall/{CRC}},\ \bibinfo {year} {2009})\BibitemShut {NoStop}%
	\bibitem [{\citenamefont {Bohnet}\ \emph {et~al.}(2012)\citenamefont {Bohnet}, \citenamefont {Chen}, \citenamefont {Weiner}, \citenamefont {Meiser}, \citenamefont {Holland},\ and\ \citenamefont {Thompson}}]{Bohnet2012}%
	\BibitemOpen
	\bibfield  {author} {\bibinfo {author} {\bibfnamefont {J.~G.}\ \bibnamefont {Bohnet}}, \bibinfo {author} {\bibfnamefont {Z.}~\bibnamefont {Chen}}, \bibinfo {author} {\bibfnamefont {J.~M.}\ \bibnamefont {Weiner}}, \bibinfo {author} {\bibfnamefont {D.}~\bibnamefont {Meiser}}, \bibinfo {author} {\bibfnamefont {M.~J.}\ \bibnamefont {Holland}}, \ and\ \bibinfo {author} {\bibfnamefont {J.~K.}\ \bibnamefont {Thompson}},\ }\href {\doibase 10.1038/nature10920} {\bibfield  {journal} {\bibinfo  {journal} {Nature}\ }\textbf {\bibinfo {volume} {484}},\ \bibinfo {pages} {78} (\bibinfo {year} {2012})}\BibitemShut {NoStop}%
	\bibitem [{\citenamefont {Norcia}\ \emph {et~al.}(2016)\citenamefont {Norcia}, \citenamefont {Winchester}, \citenamefont {Cline},\ and\ \citenamefont {Thompson}}]{Norcia2016a}%
	\BibitemOpen
	\bibfield  {author} {\bibinfo {author} {\bibfnamefont {M.~A.}\ \bibnamefont {Norcia}}, \bibinfo {author} {\bibfnamefont {M.~N.}\ \bibnamefont {Winchester}}, \bibinfo {author} {\bibfnamefont {J.~R.~K.}\ \bibnamefont {Cline}}, \ and\ \bibinfo {author} {\bibfnamefont {J.~K.}\ \bibnamefont {Thompson}},\ }\href {\doibase 10.1126/sciadv.1601231} {\bibfield  {journal} {\bibinfo  {journal} {Sci. Adv.}\ }\textbf {\bibinfo {volume} {2}},\ \bibinfo {pages} {e1601231} (\bibinfo {year} {2016})}\BibitemShut {NoStop}%
	\bibitem [{\citenamefont {Dicke}(1954)}]{Dicke1954}%
	\BibitemOpen
	\bibfield  {author} {\bibinfo {author} {\bibfnamefont {R.~H.}\ \bibnamefont {Dicke}},\ }\href {\doibase 10.1103/physrev.93.99} {\bibfield  {journal} {\bibinfo  {journal} {Phys. Rev.}\ }\textbf {\bibinfo {volume} {93}},\ \bibinfo {pages} {99} (\bibinfo {year} {1954})}\BibitemShut {NoStop}%
	\bibitem [{\citenamefont {Walls}\ and\ \citenamefont {Milburn}(2008)}]{2008a}%
	\BibitemOpen
	\bibfield  {author} {\bibinfo {author} {\bibfnamefont {D.}~\bibnamefont {Walls}}\ and\ \bibinfo {author} {\bibfnamefont {G.~J.}\ \bibnamefont {Milburn}},\ }\href {\doibase 10.1007/978-3-540-28574-8} {\emph {\bibinfo {title} {Quantum Optics}}}\ (\bibinfo  {publisher} {Springer Berlin Heidelberg},\ \bibinfo {year} {2008})\BibitemShut {NoStop}%
	\bibitem [{\citenamefont {Haken}(1983)}]{Haken1983}%
	\BibitemOpen
	\bibfield  {author} {\bibinfo {author} {\bibfnamefont {H.}~\bibnamefont {Haken}},\ }\href {\doibase 10.1007/978-3-642-45556-8} {\emph {\bibinfo {title} {Laser Theory}}}\ (\bibinfo  {publisher} {Springer Berlin Heidelberg},\ \bibinfo {year} {1983})\BibitemShut {NoStop}%
	\bibitem [{\citenamefont {Abramovici}\ \emph {et~al.}(1992)\citenamefont {Abramovici}, \citenamefont {Althouse}, \citenamefont {Drever}, \citenamefont {Gursel}, \citenamefont {Kawamura}, \citenamefont {Raab}, \citenamefont {Shoemaker}, \citenamefont {Sievers}, \citenamefont {Spero}, \citenamefont {Thorne}, \citenamefont {Vogt}, \citenamefont {Weiss}, \citenamefont {Whitcomb},\ and\ \citenamefont {Zucker}}]{Abramovici1992}%
	\BibitemOpen
	\bibfield  {author} {\bibinfo {author} {\bibfnamefont {A.}~\bibnamefont {Abramovici}}, \bibinfo {author} {\bibfnamefont {W.~E.}\ \bibnamefont {Althouse}}, \bibinfo {author} {\bibfnamefont {R.~W.~P.}\ \bibnamefont {Drever}}, \bibinfo {author} {\bibfnamefont {Y.}~\bibnamefont {Gursel}}, \bibinfo {author} {\bibfnamefont {S.}~\bibnamefont {Kawamura}}, \bibinfo {author} {\bibfnamefont {F.~J.}\ \bibnamefont {Raab}}, \bibinfo {author} {\bibfnamefont {D.}~\bibnamefont {Shoemaker}}, \bibinfo {author} {\bibfnamefont {L.}~\bibnamefont {Sievers}}, \bibinfo {author} {\bibfnamefont {R.~E.}\ \bibnamefont {Spero}}, \bibinfo {author} {\bibfnamefont {K.~S.}\ \bibnamefont {Thorne}}, \bibinfo {author} {\bibfnamefont {R.~E.}\ \bibnamefont {Vogt}}, \bibinfo {author} {\bibfnamefont {R.}~\bibnamefont {Weiss}}, \bibinfo {author} {\bibfnamefont {S.~E.}\ \bibnamefont {Whitcomb}}, \ and\ \bibinfo {author} {\bibfnamefont {M.~E.}\ \bibnamefont {Zucker}},\ }\href {\doibase 10.1126/science.256.5055.325} {\bibfield  {journal} {\bibinfo
			{journal} {Science}\ }\textbf {\bibinfo {volume} {256}},\ \bibinfo {pages} {325} (\bibinfo {year} {1992})}\BibitemShut {NoStop}%
	\bibitem [{\citenamefont {Harry}\ \emph {et~al.}(2006)\citenamefont {Harry}, \citenamefont {Armandula}, \citenamefont {Black}, \citenamefont {Crooks}, \citenamefont {Cagnoli}, \citenamefont {Hough}, \citenamefont {Murray}, \citenamefont {Reid}, \citenamefont {Rowan}, \citenamefont {Sneddon}, \citenamefont {Fejer}, \citenamefont {Route},\ and\ \citenamefont {Penn}}]{Harry2006}%
	\BibitemOpen
	\bibfield  {author} {\bibinfo {author} {\bibfnamefont {G.~M.}\ \bibnamefont {Harry}}, \bibinfo {author} {\bibfnamefont {H.}~\bibnamefont {Armandula}}, \bibinfo {author} {\bibfnamefont {E.}~\bibnamefont {Black}}, \bibinfo {author} {\bibfnamefont {D.~R.~M.}\ \bibnamefont {Crooks}}, \bibinfo {author} {\bibfnamefont {G.}~\bibnamefont {Cagnoli}}, \bibinfo {author} {\bibfnamefont {J.}~\bibnamefont {Hough}}, \bibinfo {author} {\bibfnamefont {P.}~\bibnamefont {Murray}}, \bibinfo {author} {\bibfnamefont {S.}~\bibnamefont {Reid}}, \bibinfo {author} {\bibfnamefont {S.}~\bibnamefont {Rowan}}, \bibinfo {author} {\bibfnamefont {P.}~\bibnamefont {Sneddon}}, \bibinfo {author} {\bibfnamefont {M.~M.}\ \bibnamefont {Fejer}}, \bibinfo {author} {\bibfnamefont {R.}~\bibnamefont {Route}}, \ and\ \bibinfo {author} {\bibfnamefont {S.~D.}\ \bibnamefont {Penn}},\ }\href {\doibase 10.1364/ao.45.001569} {\bibfield  {journal} {\bibinfo  {journal} {Appl. Opt.}\ }\textbf {\bibinfo {volume} {45}},\ \bibinfo {pages} {1569} (\bibinfo {year}
		{2006})}\BibitemShut {NoStop}%
	\bibitem [{\citenamefont {Graham}\ \emph {et~al.}(2013)\citenamefont {Graham}, \citenamefont {Hogan}, \citenamefont {Kasevich},\ and\ \citenamefont {Rajendran}}]{Graham2013}%
	\BibitemOpen
	\bibfield  {author} {\bibinfo {author} {\bibfnamefont {P.~W.}\ \bibnamefont {Graham}}, \bibinfo {author} {\bibfnamefont {J.~M.}\ \bibnamefont {Hogan}}, \bibinfo {author} {\bibfnamefont {M.~A.}\ \bibnamefont {Kasevich}}, \ and\ \bibinfo {author} {\bibfnamefont {S.}~\bibnamefont {Rajendran}},\ }\href {\doibase 10.1103/physrevlett.110.171102} {\bibfield  {journal} {\bibinfo  {journal} {Phys. Rev. Lett.}\ }\textbf {\bibinfo {volume} {110}},\ \bibinfo {pages} {171102} (\bibinfo {year} {2013})}\BibitemShut {NoStop}%
	\bibitem [{\citenamefont {Fortier}\ \emph {et~al.}(2007)\citenamefont {Fortier}, \citenamefont {Ashby}, \citenamefont {Bergquist}, \citenamefont {Delaney}, \citenamefont {Diddams}, \citenamefont {Heavner}, \citenamefont {Hollberg}, \citenamefont {Itano}, \citenamefont {Jefferts}, \citenamefont {Kim}, \citenamefont {Levi}, \citenamefont {Lorini}, \citenamefont {Oskay}, \citenamefont {Parker}, \citenamefont {Shirley},\ and\ \citenamefont {Stalnaker}}]{Fortier2007}%
	\BibitemOpen
	\bibfield  {author} {\bibinfo {author} {\bibfnamefont {T.~M.}\ \bibnamefont {Fortier}}, \bibinfo {author} {\bibfnamefont {N.}~\bibnamefont {Ashby}}, \bibinfo {author} {\bibfnamefont {J.~C.}\ \bibnamefont {Bergquist}}, \bibinfo {author} {\bibfnamefont {M.~J.}\ \bibnamefont {Delaney}}, \bibinfo {author} {\bibfnamefont {S.~A.}\ \bibnamefont {Diddams}}, \bibinfo {author} {\bibfnamefont {T.~P.}\ \bibnamefont {Heavner}}, \bibinfo {author} {\bibfnamefont {L.}~\bibnamefont {Hollberg}}, \bibinfo {author} {\bibfnamefont {W.~M.}\ \bibnamefont {Itano}}, \bibinfo {author} {\bibfnamefont {S.~R.}\ \bibnamefont {Jefferts}}, \bibinfo {author} {\bibfnamefont {K.}~\bibnamefont {Kim}}, \bibinfo {author} {\bibfnamefont {F.}~\bibnamefont {Levi}}, \bibinfo {author} {\bibfnamefont {L.}~\bibnamefont {Lorini}}, \bibinfo {author} {\bibfnamefont {W.~H.}\ \bibnamefont {Oskay}}, \bibinfo {author} {\bibfnamefont {T.~E.}\ \bibnamefont {Parker}}, \bibinfo {author} {\bibfnamefont {J.}~\bibnamefont {Shirley}}, \ and\ \bibinfo {author}
		{\bibfnamefont {J.~E.}\ \bibnamefont {Stalnaker}},\ }\href {\doibase 10.1103/physrevlett.98.070801} {\bibfield  {journal} {\bibinfo  {journal} {Phys. Rev. Lett.}\ }\textbf {\bibinfo {volume} {98}},\ \bibinfo {pages} {070801} (\bibinfo {year} {2007})}\BibitemShut {NoStop}%
	\bibitem [{\citenamefont {Rosi}\ \emph {et~al.}(2014)\citenamefont {Rosi}, \citenamefont {Sorrentino}, \citenamefont {Cacciapuoti}, \citenamefont {Prevedelli},\ and\ \citenamefont {Tino}}]{Rosi2014}%
	\BibitemOpen
	\bibfield  {author} {\bibinfo {author} {\bibfnamefont {G.}~\bibnamefont {Rosi}}, \bibinfo {author} {\bibfnamefont {F.}~\bibnamefont {Sorrentino}}, \bibinfo {author} {\bibfnamefont {L.}~\bibnamefont {Cacciapuoti}}, \bibinfo {author} {\bibfnamefont {M.}~\bibnamefont {Prevedelli}}, \ and\ \bibinfo {author} {\bibfnamefont {G.~M.}\ \bibnamefont {Tino}},\ }\href {\doibase 10.1038/nature13433} {\bibfield  {journal} {\bibinfo  {journal} {Nature}\ }\textbf {\bibinfo {volume} {510}},\ \bibinfo {pages} {518} (\bibinfo {year} {2014})}\BibitemShut {NoStop}%
\end{thebibliography}
\end{document}